\documentclass[twocolumn,showkeys,twocolappendix]{aastex631}
\usepackage{amsmath, amssymb}
\usepackage{mathptmx}
\usepackage{verbatim}
\usepackage{graphicx}
\usepackage{textcomp}
\usepackage{latexsym}
\usepackage{multirow}
\usepackage{natbib}
\usepackage{wasysym}
\usepackage{float}
\usepackage{hyperref}
\usepackage{academicons}
\usepackage{xcolor}
\usepackage{gensymb}
\usepackage{longtable}
\usepackage{ifthenx}

\newcommand{\ltsim}{\raisebox{-.5ex}{$\;\stackrel{<}{\sim}\;$}}

\newcommand{\kms}{\ifmmode {\text{km s}}^{-1} \else km s$^{-1}$\fi}

\newcommand{\xray}{\hbox{X-ray}}
\newcommand{\aox}{$\alpha_{\rm ox}$}

\newcommand{\nh}{$N_{\rm H}$}
\newcommand{\xmm}{{\textit{XMM-Newton}}}
\newcommand{\chandra}{{\textit{Chandra}}}
\newcommand{\swift}{{\textit{Swift}}}
\newcommand{\rosat}{{\textit{ROSAT}}}

\newcommand{\ciao}{\textsc{ciao}}

\newcommand{\pimms}{\texttt{\textsc{pimms}}}

\newcommand{\pIt}{\citetalias{2014ApJ...783..116S}}

\newcommand{\pIIt}{\citetalias{2017ApJ...848...46S}}

\newcommand{\Psevent}{\citetalias{2017MNRAS.471.4398P}}

\newcommand{\pIandII}{Papers\hyperlink{cite.2014ApJ...783..116S}{~I}~and~\hyperlink{cite.2017ApJ...848...46S}{II}}

\defcitealias{2014ApJ...783..116S}{Paper~I}
\defcitealias{2017ApJ...848...46S}{Paper~II}
\defcitealias{2017MNRAS.471.4398P}{P17}

\graphicspath{{./}{Figures/}}
\newcommand{\orcid}[1]{\href{https://orcid.org/#1}{\textcolor[HTML]{A6CE39}{\includegraphics[]{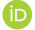}}}}

\accepted{2021 October 7}
\received{2021 July 5}
\shortauthors{Thomas et al.}
\shorttitle{\xray\ Monitoring of Luminous RQQs at High Redshift}
\begin{document}

\title{\textbf{Exploratory X-Ray Monitoring of Luminous Radio-Quiet Quasars at High Redshift:\\Extended Time-Series Analyses and Stacked Imaging Spectroscopy}}

\author[0000-0002-2456-3209]{Marcus~O.~Thomas}
\affiliation{Department of Physics, University of North Texas, Denton, TX 76203, USA;  {\rm \href{mailto:marcusthomas@my.unt.edu}{MarcusThomas@my.unt.edu}}}
\author[0000-0003-4327-1460]{Ohad~Shemmer}
\affiliation{Department of Physics, University of North Texas, Denton, TX 76203, USA;  {\rm \href{mailto:marcusthomas@my.unt.edu}{MarcusThomas@my.unt.edu}}}
\author[0000-0002-0167-2453]{W.~N.~Brandt}
\affiliation{Department of Astronomy and Astrophysics, 525 Davey Lab, The Pennsylvania State University, University Park, PA 16802, USA}
\affiliation{Institute for Gravitation and the Cosmos, The Pennsylvania State University, University Park, PA 16802, USA}
\affiliation{Department of Physics, 104 Davey Lab, The Pennsylvania State University, University Park, PA 16802, USA}
\author[0000-0003-4210-7693]{Maurizio~Paolillo}
\affiliation{Dipartimento di Scienze Fisiche, Universit\`{a} Federico II di Napoli, via Cintia 6, I-80126 Napoli, Italy}
\affiliation{Agenzia Spaziale Italiana - Science Data Center, Via del Politecnico snc, I-00133 Roma, Italy}
\affiliation{INFN - Unit\`{a} di Napoli, via Cintia 9, I-80126, Napoli, Italy}
\author[0000-0002-9925-534X]{Shai~Kaspi}
\affiliation{School of Physics \& Astronomy and the Wise Observatory, Tel Aviv University, Tel Aviv 69978, Israel}
\author[0000-0002-8853-9611]{Cristian~Vignali}
\affiliation{Dipartimento di Fisica e Astronomia ``Augusto Righi'', Universit\`{a} degli Studi di Bologna, Via Gobetti 93/2, I-40129 Bologna, Italy}
\affiliation{INAF -- Osservatorio di Astrofisica e Scienza dello Spazio di Bologna, Via Gobetti 93/3, I-40129 Bologna, Italy}
\author[0000-0003-1523-9164]{Paulina~Lira}
\affiliation{Departamento de Astronomia, Universidad de Chile, Camino del Observatorio 1515, Santiago, Chile}
\author[0000-0001-7240-7449]{Donald~P.~Schneider}
\affiliation{Department of Astronomy and Astrophysics, 525 Davey Lab, The Pennsylvania State University, University Park, PA 16802, USA}
\affiliation{Institute for Gravitation and the Cosmos, The Pennsylvania State University, University Park, PA 16802, USA}

\begin{abstract}
\noindent

 We present three new \chandra\ \xray\ epochs along with new ground-based optical-UV observations as the third installment in a time-series analysis of four high-redshift ($z \approx$ 4.1--4.4) radio-quiet quasars (RQQs). 
In total, we present nine epochs for these sources with rest-frame temporal baselines of $\sim$ 1300--2000 days. 
We utilize the X-ray data to determine basic variability properties, as well as produce mean spectra and stacked images based on effective exposure times of $\sim$40--70~ks per source. 
We perform time-series analyses in the soft and hard bands, separately, and compare variability properties to those of sources at lower redshifts and luminosities.
The magnitude of \xray\ variability of our sources remains consistent with or lower than that of similar sources at lower redshifts, in agreement with the variability-luminosity anti-correlation. 
The mean power-law photon indices in the stacked \chandra\ spectra of our sources are consistent with the values measured from their archival \xmm\ spectra separated by about three years in the rest frame.
Along with the \xray\ observations we provide near-simultaneous optical monitoring of the sources in the optical-UV regime. 
The overall variability in the optical-to-\xray\ spectral slope is consistent with sources at lower redshifts and the optical-UV observations display mild variability on monthly timescales.
\end{abstract}
\section{Introduction}
\label{sec:intro}

Long-term variability studies of distant quasars are important for understanding the evolution of active galactic nuclei (AGN) over cosmic time \citep[e.g.,][]{2007ApJ...659..997K, 2008A&A...488...73T, 2012ApJ...753..106M, 2016ApJ...831..145Y, 2020MNRAS.498.4033T}.
\xray s, in particular, are critical to study as they generally have stronger variations on shorter timescales than those of lower energy bands \citep[see, e.g.,][]{1993ARA&A..31..717M}. 
\xray\ variability studies can provide powerful probes of quasars' inner $\sim$10  gravitational radii and important information on the structure of the central engine \citep[see, e.g.,][]{1993ApJ...414L..85L, 2014ApJ...781..105L, 2020ApJ...898L...1R}, including the mass and accretion rate of the supermassive black hole \citep[see, e.g.,][hereafter, P17]{2006Natur.444..730M, 2013ApJ...779..187K, 2014ApJ...787L..12L,2017MNRAS.471.4398P}.

Nearby quasars show an anti-correlation between \xray\ variability and luminosity \citep[e.g.,][]{1993ApJ...414L..85L}.
However, their high-redshift counterparts are poorly sampled (see, e.g., \citealt{2017A&A...603A.128N}, \citealt{2019A&A...630A.118V}, and  \citealt{2021ApJ...908...53W} for recent reviews), and this trend was not so clear for sources at $z\gtrsim1$ \citep[see, e.g., ][]{2011A&A...536A..84V, 2012ApJ...746...54G,2016ApJ...831..145Y}, including the peak of quasar activity \citep[e.g.,][]{2006AJ....131.2766R}.
Quite a few high-redshift, and therefore luminous, AGNs have displayed significant \xray\ variability on timescales of months to years (see, e.g., \citealt{2004ApJ...611...93P}; \citealt[][hereafter, Paper~I]{2014ApJ...783..116S}; 
\citealt{2016A&A...593A..55V}). Several studies have suggested that the observed variability evolves with redshift \citep[e.g.,][]{2000MNRAS.315..325A,2002MNRAS.330..390M,2004ApJ...611...93P,2008A&A...487..475P}, which is difficult to reconcile with the relative consistency of the basic \xray\ properties of AGNs up to $z\sim7$ \citep[e.g., ][]{2005ApJ...630..729S,2006ApJ...646L..29S,2005AJ....129.2519V,2006AJ....131.2826S,2007ApJ...665.1004J,2019A&A...630A.118V}. 
Other works suggest, instead, a lack of evolution with redshift (e.g., \citealt{2014ApJ...781..105L}; \Psevent).

In light of this discrepancy, we began an \xray\ monitoring survey using the \chandra\ \xray\ \textit{Observatory} \citep[hereafter, \chandra;][]{2000SPIE.4012....2W} of four representative radio-quiet quasars 
(RQQs\footnote{Radio-quiet AGNs are defined as sources with \hbox{$R = f_\nu (5~{\text{GHz}})/f_\nu$(4400~\AA) $< 10$}, where $f_\nu(5~{\text{GHz}})$ and $f_\nu$(4400~\AA) represent flux densities at rest-frame 5~GHz and 4400~\AA, respectively (\citealt{1989AJ.....98.1195K}). 
Radio-quiet objects are selected to minimize potential interference by jet-related \xray\ variations on the accretion-disk-corona system.})
with redshifts between 4.10 and 4.35 (hereafter ``\chandra\ sources''; see \pIt).
The sources, \hbox{Q $0000-263$}, \hbox{BR $0351-1034$}, \hbox{PSS $0926+3055$}, and \hbox{PSS $1326+0743$}, were selected out of the sample of \cite{2005ApJ...630..729S} as having the highest accessible redshifts for economical Chandra observations and having at least two archival \xray\ epochs.

\pIt\ presented the variability information based on four \xray\ epochs, two of which were archival, which was compared to a sample of similarly luminous quasars at considerably lower redshifts ($1.33<z<2.74$), observed with the \textit{The Neil Gehrels Swift Observatory} (\citealt{2004ApJ...611.1005G}; hereafter ``\swift\ sources''), as well as nearby quasars (i.e., \hbox{$0\lesssim z \lesssim 0.3$}) from \cite{1998ApJ...503..607F}. 
The initial findings revealed that the \chandra\ sources exhibited \xray\ variability similar to far less-luminous nearby AGN, which appeared at odds with the anti-correlation between \xray\ variability and luminosity observed in the nearby universe.
However, the higher-redshift \chandra\ sources exhibited, on average, a smaller variability amplitude than the \swift\ sources, indicating that \xray\ variability does not necessarily increase with redshift. Further, three of the four \chandra\ sources were considered \xray\ variable following a $\chi^2$ test on their light curves.

\citet[][hereafter, Paper~II]{2017ApJ...848...46S} strengthened \pIt's conclusion that \xray\ variability does not depend on redshift and the addition of two \chandra\ epochs for each of our \chandra\ sources reduced the number of variable objects to one.
When considering only the \chandra\ observations of the \chandra\ sources, no significant \xray\ variability was detected in any of these sources.
\pIIt\ also presented a comparison with the ``bright-R'' sample (\hbox{$0.42\leqslant z \leqslant 3.70$}; \Psevent) of 94 quasars from the 7~Ms \chandra\ Deep Field-South survey \citep[CDF-S;][]{2017ApJS..228....2L}, spanning over 17 years in the observed frame.
The comparison of the \chandra, \swift, and the less-luminous CDF-S sources revealed the trend of increasing \xray\ variability amplitude with decreasing luminosity, and offered no evidence of \xray\ variability increasing with redshift.

In this paper, we present three additional \chandra\ epochs for the \chandra\ sources (nine \xray\ epochs in total), that allow us to extend our time-series analysis and produce tighter constraints on the X-ray properties of these sources. 
Using these new data, we are now able to present mean imaging spectroscopy and deep \xray\ imaging spanning three and a half years in the rest-frame for \hbox{PSS $0926+3055$} and \hbox{PSS $1326+0743$}, and nearly two years in the rest-frame for \hbox{Q $0000-263$} and \hbox{BR $0351-1034$}.
We also perform all prior and new data reduction and variability analysis on the \chandra\ data using a more reliable method for flux calculation, extend the analysis into the \chandra\ hard band (2-8~keV), and perform Monte Carlo simulations to strengthen our variability classifications.

This paper is organized as follows:
Section \ref{sec:obs} presents the data obtained from the three new epochs, as well as our means of data reduction and analysis. 
Section \ref{sec:results} details the need for and implementation of light curve simulations for each source, extends our prior time-series analysis, and presents stacked \chandra\ images and mean spectra. 
This section also presents our ground-based optical observations of these objects taken close in time to the \chandra\ observations.
Section \ref{sec:summary} provides a summary of our findings.
Appendix \ref{apx:a} describes the effects of contamination buildup on the \chandra\ detector pertaining to our analysis, and Appendix \ref{apx:b} reports the spectral properties of previously published epochs reprocessed with the new methods used in this work.
Luminosity distances were computed using the standard cosmological model \citep[\hbox{$\Omega_{\Lambda}=0.7$}, \hbox{$\Omega_{\rm M}=0.3$}, and \hbox{$H_0=70$~\kms~Mpc$^{-1}$}; e.g.,][]{2007ApJS..170..377S}. 

\begin{deluxetable*}{lcccccllc}[ht] 
\tablecolumns{9}
\tablecaption{Log of New \chandra\ Observations}
\label{tab:chandra_log}
\tablehead{
\colhead{} &
\colhead{} &
\colhead{} &
\colhead{} &
\colhead{Galactic \nh\tablenotemark{a}} &
\colhead{} &
\colhead{} &
\colhead{} &
\colhead{Exp. Time\tablenotemark{b}} \\
\colhead{Quasar} &
\colhead{$\alpha$ (J2000.0)} &
\colhead{$\delta$ (J2000.0)} &
\colhead{$z$} &
\colhead{(10$^{20}$\,cm$^{-2}$)} &
\colhead{Cycle} &
\colhead{Obs. Date} & 
\colhead{Obs. ID} &
\colhead{(ks)}
}
\startdata
\object{Q~0000$-$263}  & 00 03 22.9 & $-$26 03 16.8 & 4.10 & 1.67 & 19 & 2018 Sep 1 & 20606 & 9.83 \\
    &\nodata & \nodata& & \nodata & 20 & 2019 Aug 13  & 20607\tablenotemark{c} & 9.85\\
    &\nodata & \nodata& & \nodata & 21 & 2020 Aug 15 & 20608 & 9.94\\
\object{BR~0351$-$1034} & 03 53 46.9 &  $-$10 25 19.0 & 4.35 & 4.08 & 19 & 2018 Oct 24 & 20609 & 9.94 \\
    &\nodata & \nodata& & \nodata & 20 & 2019 Nov 1  & 20610\tablenotemark{c} & 9.83\\
    &\nodata & \nodata& & \nodata & 21 & 2020 Oct 28 & 20611 & 9.26\\
\object{PSS~0926$+$3055} & 09 26 36.3 & $+$30 55 05.0 & 4.19 & 1.89 & 19 & 2018 Mar 9 & 20600 & 4.99 \\
    &\nodata & \nodata& & \nodata & 20 & 2019 Jan 8  & 20601\tablenotemark{c} & 4.89\\
    &\nodata & \nodata& & \nodata & 21 & 2020 Jan 25  & 20602\tablenotemark{c} & 4.89\\
\object{PSS~1326$+$0743} & 13 26 11.9 & $+$07 43 58.4 & 4.17 & 2.01 & 19 & 2018 Apr 30  & 20603 & 4.90 \\
    &\nodata & \nodata& & \nodata & 20 & 2019 Apr 30  & 20604\tablenotemark{c} & 5.05\\
    &\nodata & \nodata& & \nodata & 21 & 2020 Mar 26  & 20605\tablenotemark{c} & 4.99\\
\enddata
\tablenotetext{a}{Obtained from \cite{1990ARAA..28..215D} using the \nh\ tool at \url{http://heasarc.gsfc.nasa.gov/cgi-bin/Tools/w3nh/w3nh.pl}}
\tablenotetext{b}{Exposure time was adjusted to account for detector dead time.}
\tablenotetext{c}{Reprocessed by the CXC as part of the Repro-V Campaign \url{https://cxc.cfa.harvard.edu/cda/repro5.html}.}
\end{deluxetable*}

\begin{deluxetable*}{lccccclllc} 
\tablecolumns{10}
\tablecaption{Basic \xray\ Measurements}
\label{tab:chandra_counts}
\tablehead{ 
\colhead{} &
\colhead{} &
\multicolumn{3}{c}{Counts\tablenotemark{\small a}} \\
\cline{3-5} \\
\colhead{Quasar} &
\colhead{Cycle} &
\colhead{0.5--2~keV} & 
\colhead{2--8~keV} &
\colhead{0.5--8~keV} & 
\colhead{Band Ratio\tablenotemark{\small b}} &
\colhead{$\Gamma_{\rm eff}$\tablenotemark{\small c}} &
\colhead{Count Rate\tablenotemark{\small d}} &
\colhead{$f_{2~\rm keV}$\tablenotemark{\small c}}
}
\startdata
\object{Q~0000$-$263} & 19
 & {\phn}27.7$^{+6.3}_{-5.2}$ 
 & {\phn}8.6$^{+4.1}_{-2.9}$ 
 & {\phn}38.2$^{+7.2}_{-6.2}$ 
 & {\phn}0.30$^{+0.27}_{-0.05}$ 
 & {\phn}2.5$^{+0.7}_{-0.6}$ 
 & {\phn}2.8$^{+0.6}_{-0.5}$ 
 & 2.85 \\
& 20 
& {\phn}19.7$^{+5.5}_{-4.4}$  
& {\phn}6.8$^{+3.7}_{-2.5}$ 
& {\phn}26.5$^{+6.2}_{-5.1}${\phn}
& {\phn}0.37$^{+0.40}_{-0.06}$ 
& {\phn}1.8$^{+0.8}_{-0.7}$
& {\phn}2.0$^{+0.6}_{-0.4}$ 
& 0.97 \\
& 21 
& {\phn}23.5$^{+5.9}_{-4.8}$ 
& {\phn}18.8$^{+5.4}_{-4.3}$ 
& {\phn}42.2$^{+7.6}_{-6.5}$
& {\phn}0.78$^{+0.43}_{-0.12}$ 
& {\phn}2.0$\pm0.5$
& {\phn}2.4$^{+0.6}_{-0.5}$ 
& 2.29 \\
\object{BR~0351$-$1034} & 19 
& {\phn}14.8$^{+4.9}_{-3.8}$ 
& {\phn}3.9$^{+3.2}_{-1.9}$ 
& {\phn}18.6$^{+5.4}_{-4.3}$ 
& {\phn}0.23$^{+0.47}_{-0.03}$ 
& {\phn}1.9$\pm1.0$ 
& {\phn}1.5$^{+0.5}_{-0.4}$ 
& 0.99 \\
& 20 
& {\phn}8.4$^{+4.0}_{-2.8}$ 
& {\phn}9.8$^{+4.2}_{-3.1}$ 
& {\phn}18.2$^{+5.4}_{-4.2}$ 
& {\phn}1.19$^{+1.76}_{-0.22}$ 
& {\phn}1.4$\pm0.8$ 
& {\phn}0.9$^{+0.4}_{-0.3}$ 
& 0.48\\
& 21 
& {\phn}7.9$^{+3.9}_{-2.7}$ 
& {\phn}5.0$^{+3.4}_{-2.1}$ 
& {\phn}12.8$^{+4.7}_{-3.5}$ 
& {\phn}0.56$^{+1.35}_{-0.09}$ 
& {\phn}2.4$^{+1.3}_{-1.1}$ 
& {\phn}0.8$^{+0.4}_{-0.3}$ 
& 0.78 \\
\object{PSS~0926$+$3055} 
& 19 
& {\phn}8.9$^{+4.1}_{-2.9}$ 
& {\phn}8.9$^{+4.1}_{-2.9}$ 
& {\phn}18.8$^{+5.4}_{-4.3}$ 
& {\phn}0.95$^{+1.31}_{-0.18}$ 
& {\phn}1.5$^{+1.1}_{-1.1}$ 
& {\phn}1.8$^{+0.8}_{-0.6}$ 
& 0.67 \\
& 20 
& {\phn}25.8$^{+5.7}_{-4.6}$ 
& {\phn}10.8$^{+4.4}_{-3.2}$ 
& {\phn}37.5$^{+7.1}_{-6.0}$ 
& {\phn}0.41$^{+0.32}_{-0.06}$ 
& {\phn}2.1$\pm0.6$ 
& {\phn}5.5$^{+1.3}_{-1.1}$ 
& 4.28\\
& 21 
& {\phn}16.5$^{+5.2}_{-4.0}$ 
& {\phn}12.6$^{+4.7}_{-3.5}$ 
& {\phn}30.1$^{+6.6}_{-5.5}$ 
& {\phn}0.74$^{+0.55}_{-0.13}$ 
& {\phn}1.4$\pm$0.8 
& {\phn}3.4$^{+1.0}_{-0.8}$ 
& 1.16 \\
\object{PSS~1326$+$0743} 
& 19 
& {\phn}18.6$^{+5.4}_{-4.3}$ 
& {\phn}4.9$^{+3.4}_{-2.1}$ 
& {\phn}28.3$^{+6.4}_{-5.3}$ 
& {\phn}0.23$^{+0.37}_{-0.03}$ 
& {\phn}1.7$^{+0.6}_{-0.7}$ 
& {\phn}3.8$^{+1.1}_{-0.9}$ 
& 1.46 \\
& 20 
& {\phn}21.7$^{+5.7}_{-4.6}$ 
& {\phn}20.7$^{+5.6}_{-4.5}$ 
& {\phn}42.3$^{+7.6}_{-6.5}$ 
& {\phn}0.94$^{+0.56}_{-0.14}$ 
& {\phn}1.6$\pm$0.5 
& {\phn}4.3$^{+1.1}_{-0.9}$ 
& 2.14 \\
& 21 
& {\phn}15.8$^{+5.1}_{-3.9}$ 
& {\phn}9.8$^{+4.2}_{-3.1}$ 
& {\phn}25.5$^{+6.1}_{-5.0}$ 
& {\phn}0.59$^{+0.59}_{-0.10}$ 
& {\phn}1.8$\pm0.8$ 
& {\phn}3.2$^{+1.0}_{-0.8}$ 
& 2.21 \\
\enddata
\tablenotetext{a}{The 1$\sigma$-level count errors were computed using Tables 1 and 2 of \citet{1986ApJ...303..336G} with Poisson statistics. Note that 1$\sigma$ corresponds to $\sim$84\% confidence in the Poisson limit.
Counts at energies below $\sim0.5$~keV have suffered significant QE losses and were omitted from this table due to an upper limit of three counts per source per Cycle (see Appendix \ref{apx:a}).}
\tablenotetext{b}{Ratio of hard- to soft-band counts. To avoid the failure of the standard approximate-variance formula at small counts, band ratio and its 1-$\sigma$ level upper and lower limits were calculated using the software Bayesian Estimation of Hardness Ratios \citep[\texttt{\textsc{behr}},][]{2006ApJ...652..610P}.}%
\tablenotetext{c}{Effective photon indices (0.5--8.0~keV; 90\% confidence) were estimated by spectral modeling in \texttt{Sherpa}. See Section \ref{sec:obs} for details on the assumed model.
Galactic absorption-corrected flux densities at rest-frame 2~keV in units of $10^{-31}$~erg~cm$^{-2}$~s$^{-1}$~Hz$^{-1}$ were estimated with the \pimms\ command-line tool v4.11b \citep{1993Legac...3...21M} by extrapolating the respective soft-band flux (see Table \ref{tab:lc_chandra}) and $\Gamma_{\rm eff}$.}
\tablenotetext{d}{Count rate computed in the soft band (observed-frame \hbox{0.5--2~keV}) in units of $10^{-3}$~counts~s$^{-1}$.}
\tablecomments{See Appendix \ref{apx:b} for updated measurements, using the new reduction methods, from epochs reported in \pIandII.}
\end{deluxetable*}


\section{X-RAY Observations and Data Reduction} \label{sec:obs}

\pIIt\ reported on six epochs from each of the four sources gathered by \chandra\ between 2003 and 2017, along with observations from \xmm\ \citep{2001A&A...365L...1J} from 2002 and 2004, and \textit{ROSAT} \citep{1981SSRv...30..569A} observations from 1991 and 1992. 
Three new epochs are provided here, per source, all obtained with \chandra\ Advanced CCD Imaging Spectrometer \citep[ACIS;][]{2003SPIE.4851...28G} snapshots during Cycles 19, 20, and 21. 
The \chandra\ observation log is presented in Table \ref{tab:chandra_log}.
The data reduction methods used in \pIandII\ have been shown to potentially produce significant bias in low-count \xray\ observations due to older and out-of-date instrument calibrations\footnote{See \url{https://cxc.harvard.edu/ciao/why/pimms.html}.}, and have therefore been retired in favor of new methods described below.
All previously reported \chandra\ observations have been reduced using these new tools, which consider regularly updated quantum efficiency (QE) degradation and charge transfer inefficiency (CTI) models, as described in Appendix \ref{apx:a}, and their spectral properties are reported in Appendix \ref{apx:b}.

Source counts and fluxes were extracted using \chandra\ Interactive Analysis of Observations
({\sc{ciao})\footnote{\url{http://cxc.cfa.harvard.edu/ciao/}} v4.12} tools and {\ciao}\ Calibration Database ({\sc{caldb}})\footnote{\url{http://cxc.harvard.edu/ciao/ahelp/caldb.html}} v4.9.3. 
All \chandra\ observations were reprocessed with the \hbox{\textit{chandra\textunderscore repro}} script to apply the latest instrument calibrations. 
In each case, the source-count results were consistent with those previously reported. 

\hyperlink{P1}{Papers~I}~and~\hyperlink{P2}{II} reported on the ultrasoft ($0.3-0.5$~keV) band, but given the lack of detectable counts (amounting to a formal upper limit of three counts) from each object in the new epochs, this band is omitted in this work.
Observed-frame counts in the \xray\ \hbox{soft (0.5-2~keV)}, \hbox{hard (2-8~keV)}, and \hbox{full (0.5-8~keV)} bands were measured using \texttt{wavdetect} \citep{2002ApJS..138..185F}, utilizing wavelet transforms with a false positive threshold of $10^{-3}$, as the \xray\ coordinates were known a priori, and were confirmed by visual inspection of the images. 
For the new epochs, Table \ref{tab:chandra_counts} reports the counts in each band, the soft-band count rate, effective power-law photon index ($\Gamma_{\text{eff}},\ 0.5-8.0~{\rm keV}$), band ratio, calculated with \texttt{Bayesian Estimation of Hardness Ratios} \citep[\textsc{\texttt{behr;}}][]{2006ApJ...652..610P}, and Galactic absorption-corrected flux density at rest-frame 2~keV. 
Effective photon indices and their $90\%$ uncertainties were estimated using spectral modeling in \texttt{Sherpa} \citep{2001SPIE.4477...76F}.
The flux-density at rest-frame 2-keV measurement methods are described below.

Soft- and hard-band unabsorbed source fluxes were obtained using two methods.
Previously, \pIandII\ reported fluxes obtained using \chandra~\textsc{pimms}\footnote{\url{https://cxc.harvard.edu/toolkit/pimms.jsp}}, which is purposed for observation planning and estimates a model-based flux with user-provided parameters. 
However, while popular, this method has pitfalls when used to calculate fluxes for observed data.
The tool is intended to function as a proposal-planning toolkit, and the calibrations applied to the latest \chandra\ Cycles are \textit{predictions} rather than based on up-to-date instrument conditions. 
While \chandra~\textsc{pimms} considers these calibration factors, the models are predictions of what the instrument conditions will be at the beginning of the next observation Cycle, $\sim$18 months in the future, and are not updated beyond the initial prediction.

According to the \chandra\ \xray\ Center (CXC), due to the loss of sensitivity to lower-energy photons, as well as the loss of effective area due to contamination buildup, the most reliable way to calculate fluxes is to use the \textsc{ciao} thread \textit{srcflux} on observed data.
The \textit{srcflux} thread is a wrapper that calculates both the model-independent position-based flux using the thread \textit{eff2evt}, as well as a power-law with photoelectric absorption model-based flux with \textit{modelflux}, given user-provided photon indices and $N_H$ parameters.  
Using this thread, we extracted unabsorbed source fluxes in the soft and hard bands, accounting for up-to-date CCD effective-area loss, CTI, and point spread function (PSF) contribution. 

When specifying an energy range for \textit{srcflux} to integrate over, the user must provide a monochromatic energy to properly weight the detector response area based on the PSF contribution. 
We used the monochromatic energy 1.35~keV and 3.4~keV in the soft and hard band, respectively, based on the absorption edges in the response near the band boundaries.
With the loss of effective area over the range of this monitoring program, the selection of a single monochromatic energy over all epochs can cause significant offsets in flux\footnote{\url{https://cxc.harvard.edu/ciao/why/monochromatic_energy.html}}.
Further, the user-provided photon index for each object, in this case, was the photon index of its mean spectrum over all epochs (see Section \ref{sec:meanspectra}).
Given that our analysis is contingent on the behavior of the sources at each individual epoch, more care should be taken when modeling the independent observations. 

In order to find the most accurate model-based flux measurement, in addition to the \textit{srcflux} estimates, the source spectrum of each \chandra\ observation was extracted from the level-2 event file with the thread \textit{specextract}. 
The spectra were fit in \texttt{Sherpa} v14.3 using the \texttt{XSPEC} \citep{1996ASPC..101...17A} \texttt{xspowerlaw} single power-law and \texttt{xsphabs} Galactic-absorption models, along with a free normalization parameter.
The \texttt{xsphabs} Galactic-absorption column density parameter for each source was frozen to the values in Table \ref{tab:chandra_log}.
The fits were made using the \textit{cstat} statistic  over the \hbox{0.5--8.0~keV} energy range with the data grouped into a minimum of one count per energy bin \citep[e.g.,][see Section \ref{sec:sims} on $\chi^2$ statistics in this low-count regime]{2017AA...605A..51K}.
Typically, it is recommended that the background spectrum be modeled along with the source spectrum when using the Cash statistic \citep{1979ApJ...228..939C}.
However, all of our observations are on-axis and background counts in the 2\arcsec\ source regions are negligible (i.e., $\langle N_{\textsc{bg}}\rangle\ll1$). 
Once the best-fit model was determined, all parameters were frozen and the ``convolution'' model component \texttt{xscflux} was applied to the power-law model component to calculate the unabsorbed energy flux to $90\%$ confidence.
The rest-frame unabsorbed 2-keV flux densities were estimated by extrapolating $\Gamma_{\rm eff}$ and the respective soft-band flux using the \pimms\ command-line tool v4.11b \citep{1993Legac...3...21M}. 

To summarize, the soft- and hard-band source fluxes were estimated in two ways: (1) \textit{srcflux}, with a single mean photon index for all epochs, per source, and a  constant monochromatic energy for the soft and hard band, respectively, and (2) by manual spectral fitting in \texttt{Sherpa} without assumed photon indices. 
Both of these methods avoid the potential offset caused by \chandra~\textsc{pimms} predictions.
Individual spectral modeling was favored over \textit{srcflux} in all proceeding analyses in light of overall smaller uncertainties.
However, we retain results from both methods for the following reasons: (1) to observe if the selection of a single $\langle\Gamma\rangle$ and monochromatic flux produced a significant effect on our sources' flux estimates, (2) for comparison with the previously published results, and (3) for reference about the aforementioned parameter-selection effects on both methods of calculating flux in low-count observations for future studies.
The results from both methods are reported in Table \ref{tab:lc_chandra}, along with those of the archival \xmm\ and \rosat\ observations.

\startlongtable
\begin{deluxetable*}{lccccccl} 
\tabletypesize{\footnotesize}
\tablecolumns{8}
\tablecaption{Long-Term X-ray Light Curve Data}
\tablehead{
\colhead{} &
\colhead{} &
\multicolumn{4}{c}{Flux ($10^{-15}$~erg~cm$^{-2}$~s$^{-1}$)}\\
\cline{3-6}\\
\colhead{Quasar} &
\colhead{JD} &
\multicolumn{2}{c}{$f_{\rm 0.5 - 2~keV}$\tablenotemark{\small \dag}} & 
\multicolumn{2}{c}{$f_{\rm 2 - 8~keV}$\tablenotemark{\small \dag}} &
\colhead{Observatory} &
\colhead{Reference}
}
\startdata
\object{Q~0000$-$263} 
& 2448588.5 & 22$\pm$3 	& \nodata	& \nodata & \nodata 	& \rosat 	& 1, 2, 3 \\
& 2452450.5 & 13$\pm$1 	& \nodata	& \nodata & \nodata 	& \xmm\ 	& 4, 5, 6 \\
& 2455802.5 & 27	$^{+6 }_{-5 }$ &	27	$^{+6 }_{-6 }$	 & 	25	$^{+5 }_{-5 }$ &	28	$^{+14}_{-6 }$	 & \chandra\ & 7 \\
& 2456173.5 & 23	$^{+5 }_{-4 }$ &	23	$^{+6 }_{-6 }$	 & 	32	$^{+7 }_{-6 }$ &	36	$^{+16}_{-6 }$	 & \chandra\ & 7 \\
& 2456540.5 & 18	$^{+5 }_{-4 }$ &	18	$^{+5 }_{-6 }$	 & 	21	$^{+6 }_{-5 }$ &	23	$^{+9 }_{-6 }$	 & \chandra\ & 8 \\
& 2456917.0 & 25	$^{+6 }_{-5 }$ &	27	$^{+8 }_{-8 }$	 & 	30	$^{+7 }_{-6 }$ &	24	$^{+14}_{-8 }$	 & \chandra\ & 8 \\
& 2458362.5 & 23	$^{+7 }_{-6 }$ &	28	$^{+10}_{-10}$	 & 	17	$^{+5 }_{-4 }$ &	20	$^{+13}_{-10}$	 & \chandra\ & 9 \\
& 2458708.5 & 16	$^{+6 }_{-5 }$ &	21	$^{+9 }_{-9 }$	 & 	24	$^{+9 }_{-7 }$ &	14	$^{+11}_{-9 }$	 & \chandra\ & 9 \\
& 2459076.5 & 30	$^{+8 }_{-7 }$ &	28	$^{+11}_{-11}$	 & 	30	$^{+8 }_{-7 }$ &	38	$^{+17}_{-11}$	 & \chandra\ & 9 \\
\newline\\
\cline{2-2}
& Mean Flux	& 22	$\pm2$	       & 	23	$\pm2$  		 &	26	$\pm2$		   &	26	$\pm3$			& \nodata\\
\cline{2-7}
\object{BR~0351$-$1034}
& 2448647.5 & 15$\pm$6 			&  \nodata 		&  \nodata & \nodata  	& \rosat 	& 2, 3 \\
& 2453035.5 & 12$\pm$2 			&  \nodata 		& \nodata  & \nodata  	& \xmm\ 	& 5, 6, 10 \\
&2455827.5 & 6	$^{+3 }_{-2 }$ &	6	$^{+4 }_{-3 }$	 & 	8	$^{+4 }_{-3 }$ &	6	$^{+8 }_{-3 }$	 & \chandra\ & 7 \\
&2455862.5 & 4	$^{+2 }_{-2 }$ &	5	$^{+3 }_{-2 }$	 & 	9	$^{+5 }_{-4 }$ &	7	$^{+9 }_{-2 }$	 & \chandra\ & 7 \\
&2456491.5 & 12	$^{+4 }_{-4 }$ &	12	$^{+5 }_{-4 }$	 & 	10	$^{+4 }_{-3 }$ &	12	$^{+10}_{-4 }$	 & \chandra\ & 8 \\
&2456987.5 & 9	$^{+3 }_{-3 }$ &	10	$^{+5 }_{-4 }$	 & 	33	$^{+12}_{-10}$ &	20	$^{+12}_{-4 }$	 & \chandra\ & 8 \\
&2458415.5 & 14	$^{+6 }_{-4 }$ &	16	$^{+8 }_{-6 }$	 & 	18	$^{+7 }_{-6 }$ &	10	$^{+10}_{-6 }$	 & \chandra\ & 9 \\
&2458792.5 & 11	$^{+5 }_{-4 }$ &	12	$^{+7 }_{-5 }$	 & 	26	$^{+10}_{-8 }$ &	22	$^{+14}_{-5 }$	 & \chandra\ & 9 \\
&2459151.0 & 7	$^{+4 }_{-3 }$ &	9	$^{+7 }_{-5 }$	 & 	8	$^{+4 }_{-3 }$ &	10	$^{+10}_{-5 }$	 & \chandra\ & 9 \\
\newline\\
\cline{2-2}
& Mean Flux		& 10 $\pm1$	   &	11	$\pm2$			 &	16	$\pm4$		   &	12	$\pm2$			 &\nodata	\\			
\cline{2-7}
\object{PSS~0926$+$3055} 
& 2452344.5 & 33	$^{+8 }_{-6 }$ &	33	$^{+9 }_{-7 }$	 & 	52	$^{+12}_{-10}$ &	53	$^{+27}_{-7 }$	 	&\chandra\ 		& 5, 11 \\
& 2453322.5 & 39$\pm$3 		& \nodata &  \nodata &  \nodata   			& \xmm\ 		& 5 \\
& 2455623.5 & 31	$^{+8 }_{-7 }$ &	32	$^{+10}_{-8 }$	 & 	46	$^{+12}_{-10}$ &	42	$^{+25}_{-8 }$	  & \chandra\ & 7 \\
& 2455939.5 & 24	$^{+8 }_{-6 }$ &	23	$^{+9 }_{-7 }$	 & 	23	$^{+8 }_{-6 }$ &	31	$^{+22}_{-7 }$	  & \chandra\ & 7 \\
& 2456424.5 & 39	$^{+9 }_{-8 }$ &	47	$^{+13}_{-11}$	 & 	73	$^{+18}_{-15}$ &	60	$^{+29}_{-11}$	  & \chandra\ & 8 \\
& 2456675.5 & 39	$^{+10}_{-9 }$ &	41	$^{+12}_{-10}$	 & 	49	$^{+13}_{-11}$ &	40	$^{+25}_{-10}$	  & \chandra\ & 8 \\
& 2458186.5 & 14	$^{+6 }_{-5 }$ &	17	$^{+11}_{-7 }$	 & 	50	$^{+21}_{-16}$ &	40	$^{+25}_{-7 }$	  & \chandra\ & 9 \\
& 2458491.5 & 51	$^{+15}_{-12}$ &	52	$^{+18}_{-15}$	 & 	47	$^{+14}_{-12}$ &	49	$^{+28}_{-15}$	  & \chandra\ & 9 \\
& 2458873.5 & 30	$^{+10}_{-8 }$ &	36	$^{+17}_{-13}$	 & 	87	$^{+28}_{-23}$ &	62	$^{+31}_{-13}$	  & \chandra\ & 9 \\
\newline\\
\cline{2-2}
& Mean Flux	& 33	$\pm4$		   &	35	$\pm4$			 &	53	$\pm7$		   &	47	$\pm4$	          & \nodata	\\			
%
%
\cline{2-7}
\object{PSS~1326$+$0743} 
& 2452284.5 & 28	$^{+7 }_{-6 }$ &	28	$^{+8 }_{-6 }$	 & 	42	$^{+10}_{-8 }$ &	48	$^{+24}_{-6 }$	 & \chandra\ 	& 5, 11 \\
& 2453001.5 & 28$^{+2}_{-3}$   & \nodata &  \nodata &  \nodata		&  \xmm\ 	 	& 5 \\
& 2455627.5 & 29	$^{+8 }_{-7 }$ &	32	$^{+10}_{-8 }$	 & 	50	$^{+14}_{-11}$ &	38	$^{+24}_{-8 }$	  & \chandra\ & 7 \\
& 2456047.5 & 33	$^{+9 }_{-7 }$ &	35	$^{+11}_{-9 }$	 & 	48	$^{+13}_{-11}$ &	47	$^{+26}_{-9 }$	  & \chandra\ & 7 \\
& 2456632.5 & 31	$^{+10}_{-8 }$ &	32	$^{+11}_{-9 }$	 & 	18	$^{+6 }_{-5 }$ &	16	$^{+18}_{-9 }$	  & \chandra\ & 8 \\
& 2456729.0 & 49	$^{+12}_{-10}$ &	45	$^{+13}_{-11}$	 & 	42	$^{+10}_{-9 }$ &	56	$^{+29}_{-11}$	  & \chandra\ & 8 \\
& 2458239.0 & 26	$^{+9 }_{-7 }$ &	33	$^{+14}_{-11}$	 & 	47	$^{+16}_{-13}$ &	41	$^{+25}_{-11}$	  & \chandra\ & 9 \\
& 2458603.5 & 42	$^{+12}_{-10}$ &	44	$^{+17}_{-14}$	 & 	75	$^{+21}_{-17}$ &	86	$^{+35}_{-14}$	  & \chandra\ & 9 \\
& 2458934.5 & 35	$^{+13}_{-10}$ &	34	$^{+16}_{-12}$	 & 	46	$^{+17}_{-13}$ &	40	$^{+26}_{-12}$	  & \chandra\ & 9 \\
\newline\\
\cline{2-2}
& Mean Flux	& 33	$\pm3$		   &	35	$\pm2$			 &	46	$\pm5$		   &	46	$\pm7$		   & \nodata		
\enddata
\tablenotetext{\textit{\small \dag}}{Galactic absorption-corrected flux estimated by manual modeling in \texttt{Sherpa} (left) and using \ciao's automated \textit{srcflux} script (right).} 
\tablecomments{Mean fluxes are calculated over all epochs and observatories with uncertainties $\sigma/\sqrt{N_{\text{obs}}}$, where $\sigma$ is the standard deviation of the light curve. See Section \ref{sec:meanspectra} for mean fluxes over only \chandra\ epochs.}
\tablerefs{
(1) \citealt{1994AJ....108..374B};
(2) \citealt{2000AJ....119.2031K}; 
(3) \citealt{2001AJ....122.2143V}; 
(4) \citealt{2003AA...402..465F}; 
(5) \citealt{2005ApJ...630..729S}; 
(6) \citealt{2006AJ....131...55G}; 
(7) \hyperlink{cite.2014ApJ...783..116S}{Paper~I}; 
(8) \hyperlink{cite.2017ApJ...848...46S}{Paper~II}; 
(9) This work; 
(10) \citealt{2004AJ....127....1G}; 
(11) \citealt{2003AJ....125..418V} }
\label{tab:lc_chandra}
\end{deluxetable*}

\begin{figure*} 
\epsscale{0.6}
\gridline{\plotone{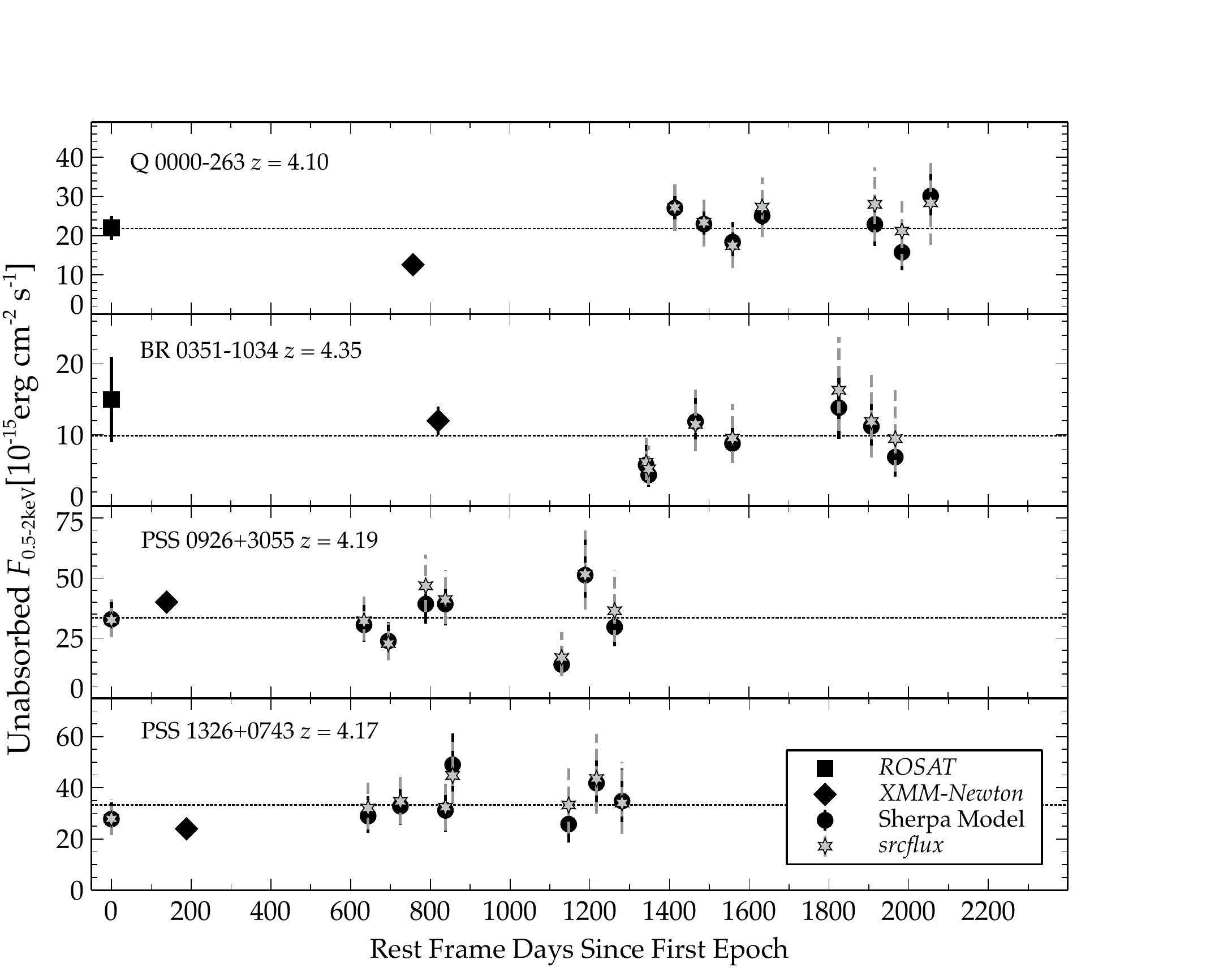}\plotone{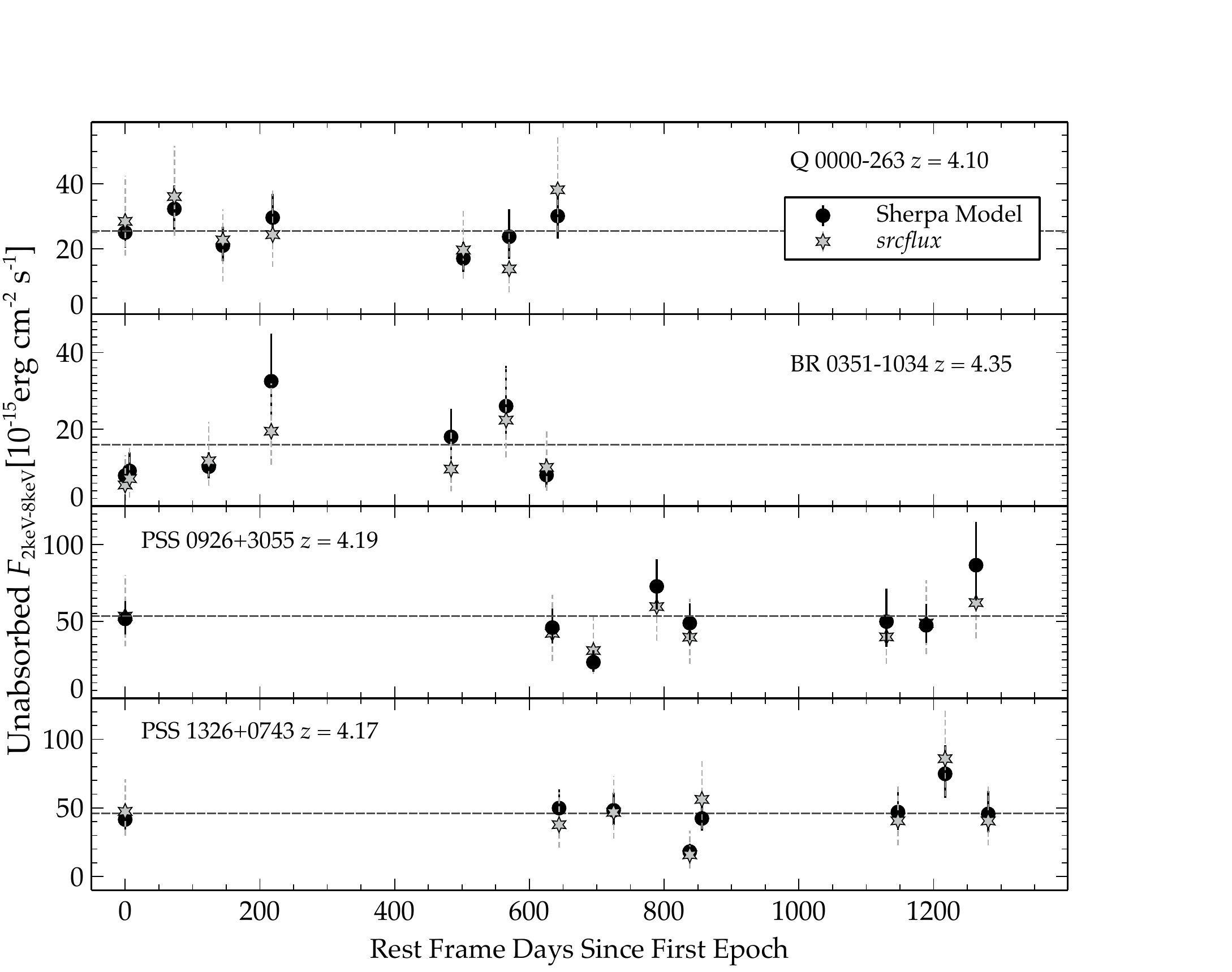}}
\caption{Galactic absorption-corrected flux in the 0.5-2 keV (left) and 2-8~keV (right) observed-frame bands plotted as a function of days in the rest frame. Squares, diamonds, circles, and grey stars represent \rosat, \xmm, \chandra\ (obtained with \texttt{Sherpa}), and \chandra\ (obtained with \textit{srcflux}) observations, respectively. 
Dotted lines represent the average flux over the light curve (using the \texttt{Sherpa} \chandra\ fluxes). 
Both methods of obtaining flux are consistent, within the uncertainties (90\%), in all cases, however, the uncertainties in the \textit{srcflux} fluxes are $\sim$33\% ($\sim$80\%) larger than the \texttt{Sherpa} fluxes in the soft (hard) band. Note that only the \chandra\ observations are shown in the hard band, and, therefore, the zero-day points differ between the soft- and hard-band light curves for \hbox{Q $0000-263$} and \hbox{BR $0351-1034$}.}
\label{fig:SBlightcurves}
\end{figure*}

\section{Results and Discussion} \label{sec:results}

The soft- and hard-band fluxes for all nine epochs of the \chandra\ sources are reported in Table \ref{tab:lc_chandra}, and the corresponding light curves are presented in Figure \ref{fig:SBlightcurves}. These light curves constitute the most detailed \xray\ light curves for RQQs at $z > 4$ and span the longest temporal baseline currently available \citep[see, e.g.,][]{2016ApJ...831..145Y}. 
As can be seen in the light curves, the resultant source fluxes using both methods are consistent within the $90\%$ uncertainties, albeit with differences in the size of the uncertainties. 
On average, the uncertainties calculated with \textit{srcflux} are $\sim$33\% larger than those obtained using \texttt{Sherpa} in the soft band, and $>80\%$ larger in the hard band, potentially due to the selection of $\langle\Gamma\rangle$ and the estimated PSF contribution from the assumed monochromatic flux (see Section \ref{sec:obs}).
Given the smaller uncertainty ranges in the \texttt{Sherpa} fluxes, all following analyses are performed on these data rather than the \textit{srcflux} measurements.

\subsection{Time-Series Analysis and Simulations}\label{sec:timeseries}

Following the prescriptions of \pIandII, we applied a $\chi^{2}$ test to a source's light curve.
In this work, we extend the test to the source's hard-band light curve, as well at their effective photon indices ($\Gamma_{\rm eff}$).
This test allows us to qualitatively determine whether a source is variable or not to a confidence of 90\% as well as make meaningful comparisons to lower redshift objects that emit at similar rest-frame energies. 
The null hypothesis in the test is that the flux (or photon index) in each epoch is consistent, within the uncertainties, with that of the mean flux (or photon index) of the object over the entire light curve. 

The $\chi^{2}$ test is defined as
\begin{equation}
	\chi^{2} = \frac{1}{N_{\rm obs}-1} \displaystyle\sum_{i=1}^{N_{\rm obs}}\frac{(f_i - \langle f \rangle)^2}{\sigma_i^2},
\end{equation}
where $f_i$ is the flux (or $\Gamma_{{\rm eff},i}$), and $\sigma_i$ its error, for the \textit{i}-th observation, $N_{\rm obs}$ is the number of epochs, and $\langle f \rangle$ is the mean flux of the light curve (or $\langle\Gamma_{\rm eff}\rangle$).
The low-count nature of our data tends to produce asymmetric uncertainty bounds. In these cases, $\sigma_i$ is the mean of the upper and lower uncertainty magnitudes. 
In order to consider effects of observatory-dependent flux calibrations, we repeated the $\chi^{2}$ test for each source using only observations taken with \chandra\ for the soft band. 
For hard-band analysis, only \chandra\ epochs were considered.

\subsubsection{$\chi^{2}$ Tests in the Low-Count Regime}\label{sec:sims}

It is well known that in the low-count regime the arrival of \xray\ events differs significantly from a Gaussian distribution, and Poisson statistics must be considered.
However, there has been debate as to the minimum number of counts per bin for the discrepancy in statistics to be negligible.
While Poisson statistics approaches a Gaussian approximation at higher event rates, the statistical bias is evident in our sample where, in several cases, less than 10 counts per band, per source, have been detected.
Additionally, the asymmetric error approximations on our flux measurements function as upper and lower limits that are larger than those approximated from a normal distribution.
This discrepancy can produce an overestimated $\chi^{2}$ statistic when assuming Gaussian uncertainties \citep[for detailed reviews on this statistical bias, see, e.g.,][]{1979ApJ...228..939C,1989ApJ...342.1207N,1999ApJ...518..380M}. 

In order to negate the potential bias and establish a reliable baseline on which to base our variability classification, we performed Monte Carlo simulations of our sources' light curves, similar to those performed in \cite{2004ApJ...611...93P}.
For each source, we simulated 1000 soft- and hard-band light curves with the assumption of non-variability over all epochs for all observatories, and once again for only \chandra\ epochs\footnote{Extending beyond $10^3$ simulated light curves does not produce a significant increase in the precision or accuracy of the simulated statistics.}.
Each simulated light curve was constructed by extracting $N_{\rm obs}$ random values from a Poisson distribution centered on the mean unabsorbed flux of the respective source.
Then, the uncertainty for each epoch in the new light curve was calculated based on the uncertainty in the observation it represented.
For example, if a source's second epoch had an uncertainty ($\sigma_i$) of 25\% of its flux, a 25\% uncertainty was applied to the second simulated observation in the light curve.

\begin{figure*}[h]
	\epsscale{1.25}
	\plotone{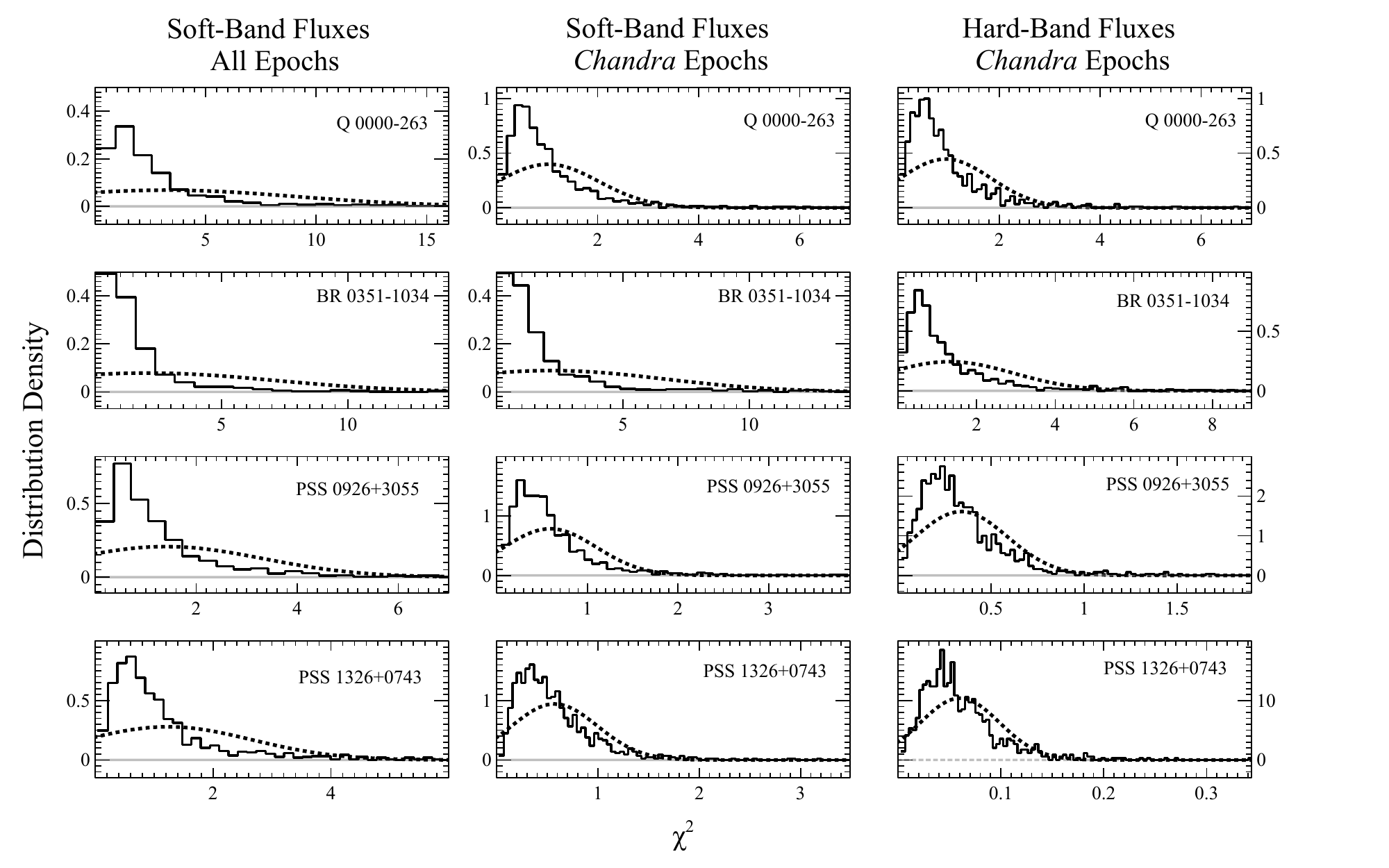}
	\caption{The results of the $\chi^2$ simulations. The solid black line in each panel plots the distribution density of $\chi^2$ statistic values from 1000 simulations of each object's light curve in the soft or hard band.
	The dotted line in each panel represents the distribution of $\chi^2$ for each set of simulations expected from a normal distribution.
		In every case, the median of the normal distribution is higher than that of the actual distribution of the simulations, suggesting a bias toward larger $\chi^2$ values in the Gaussian limit.
		Note that each plot is truncated to show at least 90\% of the actual distribution.}
\label{fig:sims}
\end{figure*}

Using these simulations, we obtained three sets of 1000 $\chi^{2}$ values for artificial \textit{non-variable} light curves of each of our four sources: (1) for all epochs in the soft band, (2) \chandra\ epochs in the soft band, and (3) \chandra\ epochs in the hard band.
The distribution densities of the simulated $\chi^2$ values are presented in Figure \ref{fig:sims}, along with the normal distribution expected from the values, and the statistical properties of each distribution are reported in Table \ref{tab:simulationStats}.
In all cases, the median of the normal distribution is higher than that of the actual distribution, showing the potential bias of higher $\chi^2$ values for our light curves in the Gaussian limit. 

To address this discrepancy, our variability classification is based on the distribution of simulated values rather than the normal distribution that the standard $\chi^{2}$ \textit{p-}value (the probability by which the null hypothesis of non-variability can be rejected) represents. 
Our \textit{p-}value based on the simulations, $p_{s}$, represents the fraction of $\chi^{2}$ in the respective source's simulated distribution that is smaller than the observed $\chi^{2}$ score. 
Therefore, an observed $\chi^{2}$ score greater than 90\% of the simulated scores (i.e., $p_s\geqslant0.90$) allows us to reject the null hypothesis of non-variability at 90\% confidence.

Table \ref{tab:variabilitySB} reports the results of the $\chi^{2}$ tests, their respective degrees of freedom (dof; where dof = $N_{\rm obs}-1$),  and $p_s$, for the soft band and hard band, respectively. 
When compared to the distribution of non-variable $\chi^{2}$ scores in the simulations, the only source for which we can reject the non-variable hypothesis over all epochs is \hbox{PSS $0926+3055$} with a score greater than 99\% of the simulated values.
 When considering only \chandra\ epochs, both \hbox{Q $0000-263$} and \hbox{PSS $0926+3055$} are variable with $p_s=0.99\text{~and~}p_s=0.90$, respectively, in the soft band.
 In the hard band, with the exception of \hbox{Q $0000-263$}, all sources are variable, and quite notably, both \hbox{PSS $0926+3055$} and \hbox{PSS $1326+0734$}'s observed $\chi^{2}$ scores were larger than 100\% of the simulated values. 
We consider all four sources variable in at least one band and \hbox{PSS $0926+3055$} in both.
 
We also performed similar Monte Carlo simulations and $\chi^2$ analysis on the $\Gamma_{\rm eff}$ values reported in Table \ref{tab:chandra_counts} (see, also,  Table \ref{tab:apxtable} for recomputed $\Gamma_{\rm eff}$ values from \pIandII).
None of the sources shows significant spectral variability in $\Gamma_{\rm eff}\ (\langle p_s\rangle=0.08$), where $\langle p_s\rangle$ is the mean $p_s$ for all four \chandra\ sources.

It should be stressed that a change in \xray\ variability classification of these sources is not necessarily an assignment of a physical secular characteristic to the source. 
Rather, it is a trend that has emerged as the light curve resolution increases, and is limited to the scope of our monitoring program. 
To say that these sources were not previously \xray\ variable, yet, are behaving so now, would be an incorrect assumption.
In fact, it has been shown that changes in variability classification should be expected as the number of net counts per source increases (see, e.g., \citealt{2012ApJ...746...54G}, \Psevent).

\subsubsection{Variability Amplitude}

To quantify the variability amplitude, \hyperlink{P1}{Papers~I}~and~\hyperlink{P2}{II} measured the excess variance, $\sigma_{rms}^2$, of each source's light curve.  
This quantity provides a measurement of the fraction of variable flux in each epoch, after subtracting statistical uncertainty. 
Both the excess variance and its uncertainty are reported in Table \ref{tab:variabilitySB}, for the soft and hard bands. 
Following \cite{1999ApJ...524..667T}, $\sigma_{rms}^2$ is defined as \
\begin{equation}
\sigma_{rms}^2 = \frac{1}{N_{obs} \langle f \rangle^2 }\displaystyle\sum_{i=1}^{N_{obs}}[(f_i - \langle f \rangle)^2 - \sigma_i^2], 
\end{equation}
where the result can be negative if the measurement errors are larger than the fraction of variable flux. 
The formal error\footnote{This expression for the error does not account for any scatter intrinsic to the red noise inherent in a typical AGN power spectral density (PSD) function (see, e.g., \citealt{2003MNRAS.345.1271V}; \citealt{2013ApJ...771....9A}).} on this value is 
$s_D/ \langle f \rangle^2\sqrt{N_{obs}}$ where, 
\begin{equation}
s_D^2 = \frac{1}{N_{obs} - 1}\displaystyle\sum_{i=1}^{N_{obs}}\{[(f_i - \langle f \rangle)^2 - \sigma_i^2] - \sigma_{rms}^2\langle f \rangle^2\}^2.
\end{equation}

The excess variance is consistent with zero, within its uncertainties, in all cases, with the exception of \hbox{BR $0351-1034$} in the hard band, where the variable flux is $\sim40\%$ of the total flux.
\hbox{BR $0351-1034$} is our faintest and highest-redshift source.
Consequently, observations of the object suffer from the lowest signal-to-noise ratio ($S/N$) of our sources, with the $\langle S/N\rangle$ being $\sim 2/3$ of that of the full sample.
Indeed, this source's hard-band $\chi^2$ score is higher than 92\% of simulations, and a non-zero result is expected; however, it should be noted that this source is subject to the highest degree of noise and statistical uncertainty among all four sources in both the soft and hard band.

Another class of results is of interest; those cases where a source is considered variable based on $p_s$, although the excess variance is consistent with zero.
This is the result in five of the six cases where $p_s\geqslant 0.90$, with the sixth being that of \hbox{BR $0351-1034$} discussed above.
The stochastic nature of AGN \xray\ emission results in aperiodic red noise dependent on the shape of the PSD function.
As a result, significant bias toward large scatter of $\sigma_{rms}^2$ that diminishes as $S/N$ increases has been shown in larger samples (see, e.g., CDF-S in Figure 4 of \Psevent).
This bias is quantifiable if a high-quality PSD is known for the source, which is not the case for our \chandra\ sources.
While the $S/N$ ratio is generally high enough for a reliable excess variance measurement, the assumed Gaussian errors and low net counts per source raises the magnitude of statistical uncertainty to that of the measured $\sigma_{rms}^2$.
Calculating the excess variance and its uncertainty in the simulated light curves shows that our low-count and sparsely sampled sources are subject to a bias toward negative values, with over 98\% of simulated $\sigma_{rms}^2 < 0$ and $\sim$25--50\% of the upper limits yet still negative in the \chandra\ epochs. 

Extreme care must be taken when using $\sigma_{rms}^2$ on individual sources.
In fact, estimates from individual light curves with low $S/N\  (\ltsim3)$ and those with sparsely sampled fluxes are highly unreliable.
Further, those estimates made on light curves with the extreme sampling patterns, such as ours, are subject to even further uncertainty \citep[see, e.g.,][]{2003MNRAS.345.1271V,2013ApJ...771....9A}.
Our sample is not ideal for $\sigma_{rms}^2$ statistics; however, we retain the results for comparison with our previously published results, based on fluxes obtained through \chandra~\textsc{pimms}, as well as other similar samples.

\begin{deluxetable*}{lcrccccrccrr} 
\label{tab:simulationStats}
\tablecolumns{12}
\tablecaption{Simulated $\chi^2$ Statistics}
\tablehead{
\colhead{} &
\multicolumn{8}{c}{Soft Band (0.5--2.0 keV)} &
\multicolumn{3}{c}{Hard Band (2.0--8.0 keV)}\\
\cline{2-8}\cline{10-12}
\colhead{} &
\multicolumn{3}{c}{All Epochs}&
\colhead{} &
\multicolumn{7}{c}{\chandra\ Epochs}\\
\cline{2-4}\cline{6-12}
\colhead{Quasar} &
\colhead{$\langle\chi^2\rangle$} &
\colhead{Min/Max} &
\colhead{$\sigma$}&
\colhead{} &
\colhead{$\langle\chi^2\rangle$} &
\colhead{Min/Max} &
\colhead{$\sigma$}&
\colhead{} &
\colhead{$\langle\chi^2\rangle$} &
\colhead{Min/Max} &
\colhead{$\sigma$}
}
\startdata
\object{Q~0000$-$263}  
 & 3.21 & 0.10 / 82.3 & 5.86 &
 & 1.02 & 0.05 / 15.1 & 1.00 &
 & 0.94 & 0.05 / 9.47 & 0.90 \\
\object{BR~0351$-$1034}  
& 2.07 & 0.07 / 77.1 & 5.06 &
& 2.15 & 0.04 / 60.6 & 4.48 &
& 1.31 & 0.03 / 19.4 & 1.62 \\
\object{PSS~0926$+$3055} 
& 1.40 & 0.03 / 33.9 & 1.92 &
& 0.59 & 0.06 / 8.35  & 0.51  & 
& 0.34 & 0.03 / 2.53 & 0.25 \\
\object{PSS~1326$+$0743} 
& 1.23 & 0.06 / 15.8 & 1.43 & 
& 0.57 & 0.03 / 4.26  & 0.42 & 
& 0.06 & $<0.01$ / 0.34 & 0.04  \\
\enddata
\tablecomments{$\sigma$ is the standard deviation of the set of simulated $\chi^2$ values. }
\end{deluxetable*}

\begin{deluxetable*}{lccrcccrcccr} 
\tablecolumns{12}
\tablecaption{\xray\ Variability Indicators}
\tablehead{
\colhead{} &
\multicolumn{8}{c}{Soft Band (0.5--2.0 keV)} &
\multicolumn{3}{c}{Hard Band (2.0--8.0 keV)}\\
\cline{2-8}\cline{10-12}
\colhead{} &
\multicolumn{3}{c}{All Epochs}&
\colhead{} &
\multicolumn{7}{c}{\chandra\ Epochs}\\
\cline{2-4}\cline{6-12}
\colhead{Quasar} &
\colhead{$\chi^2$(dof)} &
\colhead{$p_{\rm sim}$\tablenotemark{\small a}} &
\colhead{$\sigma^2_{\rm rms}$}&
\colhead{} &
\colhead{$\chi^2$(dof)} &
\colhead{$p_{\rm sim}$\tablenotemark{\small a}} &
\colhead{$\sigma^2_{\rm rms}$}&
\colhead{} &
\colhead{$\chi^2$(dof)} &
\colhead{$p_{\rm sim}$\tablenotemark{\small a}} &
\colhead{$\sigma^2_{\rm rms}$}
}
\startdata
\object{Q~0000$-$263}  
& 20.7(8)   & $0.51$ 	&  0.00$\pm$0.02  & 
& 0.77(6)	& $0.99$	& -0.02$\pm$0.02 & 
& 0.99(6) 	& $0.69$ 	& -0.02$\pm$0.02 \\
\object{BR~0351$-$1034}  
& 1.67(8)	& $0.64$ 	& -0.03$\pm$0.05 &
& 1.50(6) 	& $0.70$ 	& -0.03$\pm$0.05 &
& 3.05(6) 	& $0.92$ 	&  0.15$\pm$0.08 \\
\object{PSS~0926$+$3055} 
& 2.86(8)   & $0.99$   	& 0.03$\pm$0.04  &
& 2.40(7) 	& $0.90$	& 0.03$\pm$0.05  &
& 3.14(7) 	& $1.00$ 	& 0.03$\pm$0.05 \\
\object{PSS~1326$+$0743} 
& 1.35(8) 	& $0.74$   	&  0.00$\pm$0.02 & 
& 0.71(7) 	& $0.74$   	& -0.02$\pm$0.02 &
& 0.82(7) 	& $1.00$ 	&  0.02$\pm$0.06 \\
\enddata
\tablenotetext{a}{The fraction of simulated $\chi^2$ values smaller than the observed statistic.}
\label{tab:variabilitySB}
\end{deluxetable*}

\subsection{Variability Timescales}\label{sec:var_timescales}
\hyperlink{cite.2017ApJ...848...46S}{Paper~II} presented the first qualitative assessment of variability timescales and patterns of RQQs at $z \simeq 4.2$ by means of a variability structure function (SF).\footnote{See Section 3.3 in \hyperlink{cite.2017ApJ...848...46S}{Paper~II} on the use of SFs in the absence of a high-quality PSD function.} 
We expand on the ensemble SF presented in \hyperlink{cite.2017ApJ...848...46S}{Paper~II} by adding three additional epochs, as well as present an individual SF for each of the four \chandra\ sources, in both the soft and hard bands.
The ensemble SF was computed by averaging the SF values (i.e., $\Delta m$) of all four sources in each rest-frame time bin, according to the SF definition in \cite{1998ApJ...503..607F}, 
\begin{equation}
\Delta m_{ji} = \left|2.5\log[f(t_j)/f(t_i)]\right|,\end{equation} where $f(t_j)$ is the flux at epoch $t_j$ and $f(t_i)$ is the flux at epoch $t_i$, where $t_j>t_i$, and every $t_i$ is measured in rest-frame days since the first epoch, such that $t_i=0$. 
The uncertainty on $\langle\Delta m\rangle$ is expressed as $\sigma/\sqrt{N_{\rm b}}$ where $\sigma$ is the standard deviation of $\Delta m$ and $N_{\rm b}$ is the number of data points in the respective bin.
For the Chandra sources, we also add an additional 30\% uncertainty to $\langle\Delta m\rangle$ in quadrature to account for the \xray\ photometric errors (see Section \ref{sec:results}).

Time bins ($T_{\rm B}$) were determined using the ML-based \texttt{optbinning} Python package \citep{optbinning} and differ for each band and sample.
It should be noted that the time bins have changed from those used in \pIandII. 
We find these new time bins allow each region in the time domain to carry relatively consistent statistical weight, whereas the previously used bins were weighted in favor of longer timescales.
Each time bin now contains 7--38 data points for the \chandra\ sources.

\begin{deluxetable}{lcccl}
\tablecolumns{5}
\tablecaption{Ensemble Structure Function Time Bin Statistics}
\tablehead{
\colhead{$T_{\rm B,min}$}&
\colhead{$T_{\rm B,median}$} &
\colhead{$T_{\rm B,max}$} &
\colhead{$N_{\rm b}$\tablenotemark{\small \dag}}&
\colhead{$\langle\Delta m\rangle$}
}
\startdata
\multicolumn{5}{c}{\swift\ Sources--Soft Band}\\
\cline{1-5}
0.0		& 4.1	& 40.3		& 161	& $0.35\pm0.02$	\\
40.4	& 116.2	& 451.7		& 234	& $0.48\pm0.02$	\\
451.6	& 489.1	& 502.1		& 92	& $0.66\pm0.04$	\\
502.1	& 510.3	& 543.2		& 77	& $0.91\pm0.05$	\\
543.2 	& 733.2	& 1211.6	& 166	& $0.49\pm0.03$	\\
1211.6	& 1489.7& 1929.9	& 154	& $0.40\pm0.03$	\\
1929.9	& 2687.3& 4922.2	& 80	& $0.37\pm0.03$	\\
\multicolumn{5}{c}{\chandra\ Sources--Soft Band}\\
\cline{1-5}
6.6		& 72.6	& 138.1		& 25 	& $0.47\pm0.15$	\\
138.0	& 214.9	& 354.3		& 22	& $0.40\pm0.14$	\\
354.3	& 429.5	& 499.6		& 25	& $0.39\pm0.13$	\\
499.6	& 614.7	& 771.7		& 33	& $0.35\pm0.12$	\\
771.7	& 1000.8& 1155.5	& 18	& $0.38\pm0.13$	\\
1155.5	& 1466.2& 2056.5	& 21	& $0.41\pm0.15$	\\
\multicolumn{5}{c}{\chandra\ Sources--Hard Band}\\
\cline{1-5}
6.6		& 115.4	& 287.4		& 38 	& $0.55\pm0.18$	\\
287.4	& 351.4	& 414.0		& 13	& $0.54\pm0.22$	\\
413.9	& 429.5	& 492.7		& 13	& $0.54\pm0.19$	\\
492.8	& 503.5	& 567.4		& 12	& $0.50\pm0.21$	\\
567.4	& 639.7	& 850.4		& 15	& $0.29\pm0.12$	\\
850.3	& 1184.4& 1286.3	& 7 	& $0.29\pm0.12$	\\
\enddata
\tablenotetext{\textit{\dag}}{Number of data points in the time bin.}
\tablecomments{$T_{\rm B}$ in units of rest-frame days.}
\label{tab:eSF}
\end{deluxetable}

\begin{figure} 
\epsscale{1.2}
\plotone{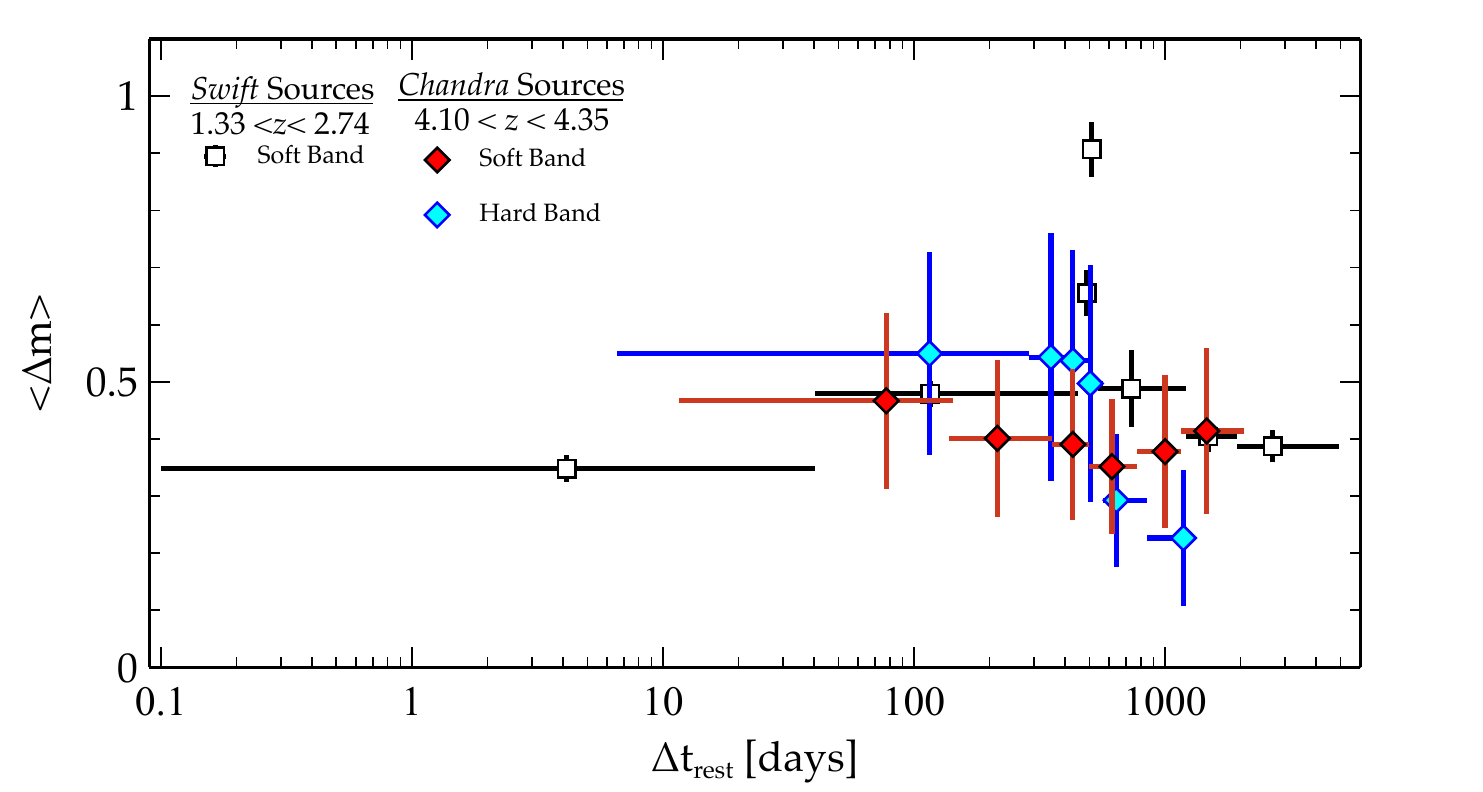}
\caption{Ensemble SFs of our \chandra\ sources, both for the hard (blue diamonds) and soft (red diamonds) bands, compared to the ensemble SF of the \swift\ sources (soft band, squares) from \hyperlink{cite.2014ApJ...783..116S}{Paper~I}. Averages of magnitude differences between all objects are plotted against time bins of rest-frame days between epochs. 
In every time bin where \swift\ and \chandra\ data are available, the soft- and hard-band \chandra\ SF is either lower than or consistent with the \swift\ SF, within the uncertainties. 
Time bins are weighted to have similar statistical weights.}  
\label{fig:ensemble}
\end{figure}

Figure \ref{fig:ensemble} presents the ensemble SF of our \chandra\ sources against the ensemble SF of the luminous \swift\ sources at intermediate redshifts \hbox{($1.33<z<2.74$)} from \hyperlink{cite.2014ApJ...783..116S}{Paper~I}.
Statistics for the time bins for the \chandra\ and \swift\ SFs are reported in Table \ref{tab:eSF}.
Given that the uncertainties on the \swift\ \xray\ photometry are considerably smaller than those of \chandra\ (see \pIt), the uncertainties on the \swift\ SF do not contain a photometric component similar to that included in the \chandra\ SFs (i.e., the addition of such a component produced a negligible effect).
In the soft and hard bands, we find that the \chandra\ SFs are consistent with or lower than that of the \swift\ SF, within the uncertainties, which is consistent with the results of \pIIt.

At all available timescales, the \chandra\ soft- and hard-band SFs are consistent within the uncertainties.
Considering the behavior of soft- and hard-band flux from both the lower- and higher-redshift sources, \xray\ variability evolution with redshift is not observed.

\begin{figure}
\centering
\epsscale{1.1}
\plotone{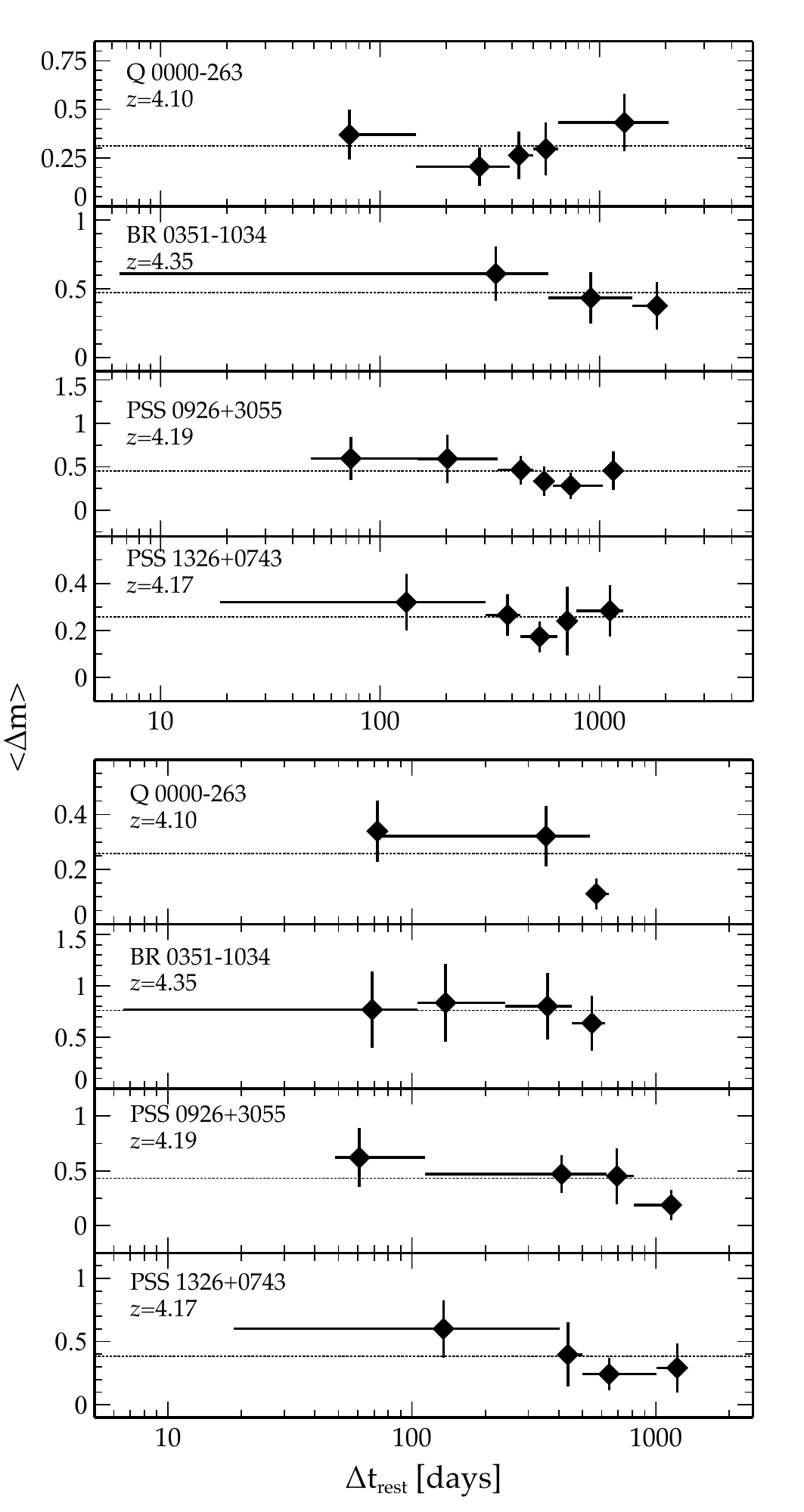}
\caption{SF of each \chandra\ object in the 0.5-2~keV band (top) and 2-8~keV band (bottom) plotted as average magnitude differences against time bins as prescribed in \cite{1998ApJ...503..607F}. 
Soft-band SFs contain data from \chandra, \xmm, and \textit{ROSAT}.
Hard-band SFs contain only \chandra\ data.
The dotted line in each panel represents the mean SF.
Time bins were constructed to hold similar statistical weights.
Within the uncertainties, nearly all points of the SFs are consistent with their respective mean in both the soft and hard bands.}
\label{fig:indvSFs}
\end{figure}

Figure \ref{fig:indvSFs} presents the SF for each \chandra\ source, in the same manner as the ensemble SF, for the soft (top) and hard (bottom) bands. 
The time bins and SF values are reported in Table \ref{tab:iSF}.
Nearly all SF values in each panel are consistent within the uncertainties, per object, at each timescale, and no clear trends of changing SF as a function of timescale are observed.
Additional and more frequent observations of the sources are needed to constrain their behavior, particularly at the shorter timescales.

\begin{deluxetable*}{lccccccccccr}
\tablecolumns{12}
\tablecaption{Individual Structure Function Time Bin Statistics}
\tablehead{
\colhead{}&
\multicolumn{5}{c}{Soft Band ($0.5-2.0$~keV)}&
\colhead{}&
\multicolumn{5}{c}{Hard Band ($2.0-8.0$~keV)}\\
\cline{2-6}\cline{8-12}
\colhead{Quasar} &
\colhead{$T_{\rm B,min}$}&
\colhead{$T_{\rm B,median}$} &
\colhead{$T_{\rm B,max}$} &
\colhead{$N_{\rm b}$\tablenotemark{\small \dag}}&
\colhead{$\Delta m$}&
\colhead{} &
\colhead{$T_{\rm B,min}$}&
\colhead{$T_{\rm B,median}$} &
\colhead{$T_{\rm B,max}$} &
\colhead{$N_{\rm b}$\tablenotemark{\small \dag}}&
\colhead{$\Delta m$}
}
\startdata
\hbox{Q $0000-263$}
&67.8	&72.6 	&145.3	&7 	& $0.37\pm0.13$& &67.8	&72.2	&73.1	&4	&$0.34\pm0.11$\\
&145.3	&283.5 	&390.4	&5  & $0.20\pm0.10$& &73.1	&354.3	&535.7	&14	&$0.32\pm0.11$\\
&390.4	&429.4 	&499.6	&5 	& $0.26\pm0.12$& &535.7	&569.8	&642.0	&3	&$0.11\pm0.06$\\
&499.6	&569.6 	&649.6	&4 	& $0.30\pm0.14$& &\nodata &\nodata &\nodata &\nodata &\nodata\\
&649.6	&1299.2 &2056.5	&15	& $0.43\pm0.15$& &\nodata &\nodata &\nodata &\nodata &\nodata\\
\cline{2-12}
\hbox{BR $0351-1034$}
&6.5	&336.8 	&584.2 	&21	& $0.61\pm0.20$& &6.5	&68.7	&105.1	&4	&	$0.77\pm0.37$\\
&684.2	&912.9 	&1407.4	&10	& $0.43\pm0.19$& &105.1	&137.5	&241.9	&5	&	$0.84\pm0.38$\\
&1407.4	&1825.8 &1963.3	&5	& $0.38\pm0.17$& &241.9	&359.6	&453.4	&5	&	$0.80\pm0.32$\\
&\nodata &\nodata &\nodata &\nodata	&\nodata & &453.4	&547.1	&621.2	&7	&	$0.64\pm0.27$\\
\cline{2-12}
\hbox{PSS $0926+3055$}
&48.4	&73.6 	&148.1 	&7 	& $0.60\pm0.24$& &48.4	&60.9	&112.9	&5	&	$0.62\pm0.27$\\
&148.1	&202.7 	&344.7 	&5 	& $0.59\pm0.28$& &112.9	&410.9	&629.0	&16	&	$0.47\pm0.17$\\
&344.7	&438.2 	&499.0 	&8 	& $0.46\pm0.17$& &629.0	&692.7	&810.3	&3	&	$0.45\pm0.25$\\
&499.0	&559.0 	&612.0 	&4 	& $0.33\pm0.17$& &810.3	&1155.0	&1258.0	&4	&	$0.19\pm0.14$\\
&612.0	&739.4 	&1032.8	&8 	& $0.28\pm0.15$& &\nodata &\nodata &\nodata &\nodata &\nodata\\
&1032.8	&1155.0 &1258.0	&4 	& $0.46\pm0.22$& &\nodata &\nodata &\nodata &\nodata &\nodata\\
\cline{2-12}
\hbox{PSS $1326+0743$}
&18.7	&131.8 	&301.3 	&11	& $0.32\pm0.12$& &18.7	&134.6	&402.5	&13	&	$0.60\pm0.23$\\
&301.3	&381.2 	&435.9 	&5 	& $0.27\pm0.09$& &402.5	&435.9	&499.7	&4	&	$0.40\pm0.25$\\
&435.9	&533.2 	&643.1 	&8 	& $0.17\pm0.07$& &499.7	&643.1	&1005.7	&8	&	$0.24\pm0.13$\\
&643.1	&711.7 	&784.4 	&4 	& $0.24\pm0.15$& &1005.7&1222.2	&1286.3	&3	&	$0.29\pm0.19$\\
&784.4	&1115.6 &1286.3	&8 	& $0.28\pm0.11$& &\nodata &\nodata &\nodata &\nodata &\nodata\\
\enddata
\tablenotetext{\textit{\dag}}{Number of data points in the time bin.}
\tablecomments{$T_{\rm B}$ in units of rest-frame days.}
\label{tab:iSF}
\end{deluxetable*}

\subsection{Stacked Images}\label{sec:deepimages}
\begin{figure*} 
\epsscale{1.2}
\plotone{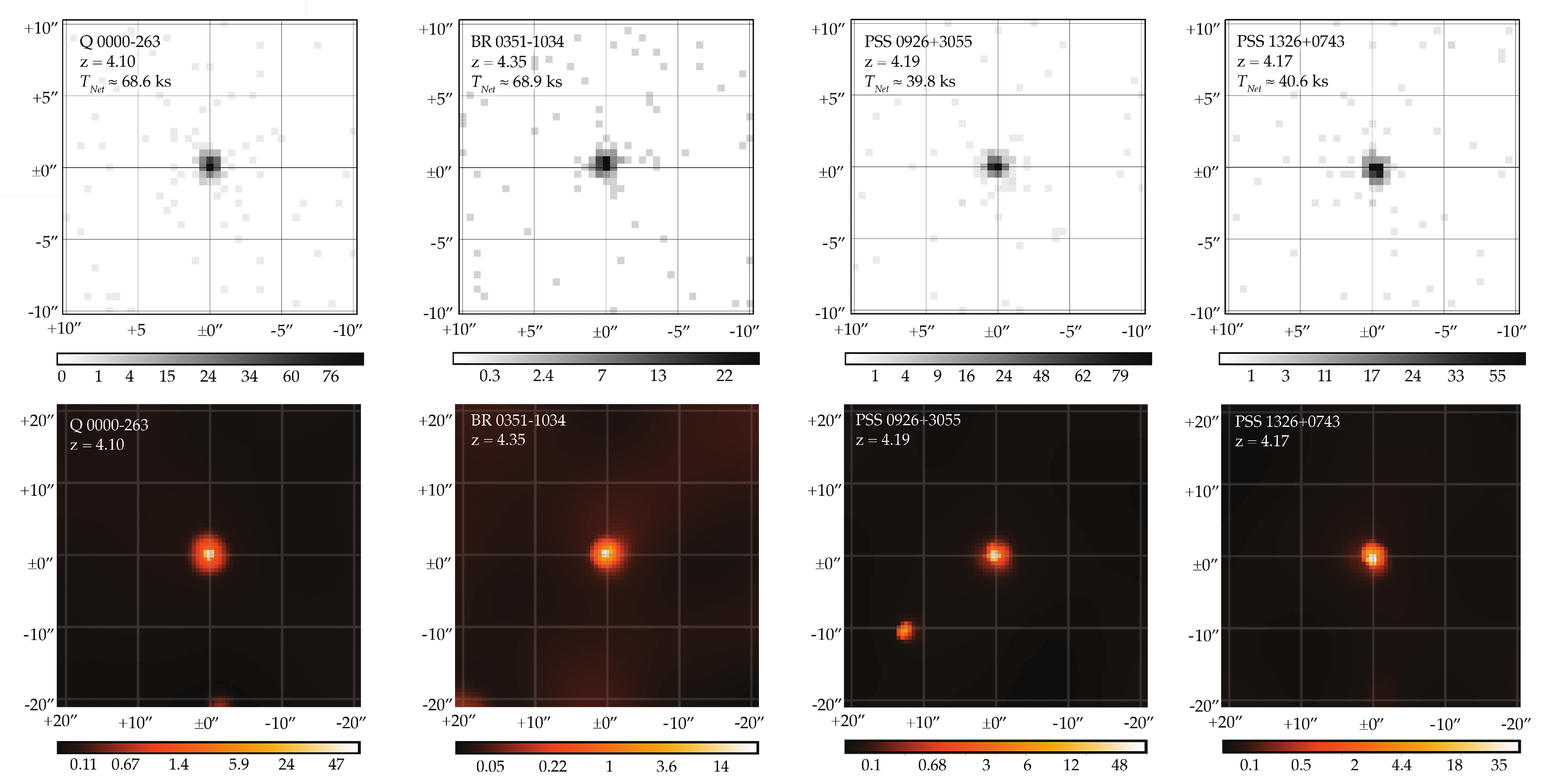}
\caption{\textbf{Top:}
Stacked, $20\arcsec\times20\arcsec$ full-band images of our four objects spanning Cycles 12-15, and 19-21, providing an effective exposure time of $\sim$70~ks for \hbox{Q $0000-263$} and \hbox{BR $0351-1034$}, and Cycles 3, 12-15, and 19-21, providing an effective exposure time of $\sim$40~ks for \hbox{PSS $0926+3055$} and \hbox{PSS $1326+0743$}. 
\textbf{Bottom:} Stacked, $40\arcsec\times40\arcsec$ full-band images of our four objects smoothed using the \textsc{ciao} thread \textit{csmooth} (see section \ref{sec:deepimages}). 
In all panels, the grid is centered on the quasar's centroid, and the color bar represents the number of counts, per pixel, by brightness.
}
\label{fig:merged} 
\end{figure*}

The stacked \chandra\ images presented in Figure \ref{fig:merged} (top) were created by first applying positional shifts to each observation.
While \chandra 's pointing accuracy is precise to $\sim 0.4\arcsec$, there are small deviations in the sources' recorded positions between observations.
For each observation, the absolute astrometry was locked to each source's position in its first observation using the \textsc{ciao} threads \textit{wcs\textunderscore match} and \textit{wcs\textunderscore update}. 
 We then used the \textit{reproject\textunderscore obs} \textsc{ciao} thread centered around each object's \xray\ coordinates to combine the observations. 
 The resulting stacked images of \hbox{PSS $0926+3055$} and \hbox{PSS $1326+0743$} contain observations from \chandra\ Cycles 3, \hbox{12-15}, and \hbox{19-21}, amounting to $\sim$40~ks each (spanning $\sim$1300 days in the rest frame), and \hbox{Q~0000-263} and \hbox{BR $0351-1034$} are composed of observations from Cycles \hbox{12-15} and \hbox{19-21}, amounting to $\sim$ 70~ks each (spanning $\sim$700 days in the rest frame). 

We utilize these images to search for extended X-ray emission or companion X-ray sources in proximity to our objects.
To aid with visual inspection of non-point like emission and companion sources, each stacked image was smoothed using the \textsc{ciao} thread \textit{csmooth}.
The images, also presented in Figure~\ref{fig:merged} (bottom), were processed using a Gaussian kernel with a minimal (maximal) signal-to-noise ratio of 2 (50), following \cite{ 2018A&A...614A.121N} for low-count observations. 
As seen in Figure \ref{fig:merged}, \hbox{BR $0351-1034$} appears to exhibit  extended emission in the southwest direction. 
However, the counts in this feature occupy $\sim$16\% of the annulus around the centroid and represent $\sim$13\% of the counts present in the total annulus, and, therefore, we consider this feature negligible.
Furthermore, there is no visible evidence for this feature in the smoothed images.  
All four sources exhibit point-like structures with no evidence of significant extended X-ray emission.

With the exception of \hbox{PSS $0926+3055$}, none of the images revealed the presence of potential companion \xray\ sources within 25\arcsec\ following a \textit{wavdetect} search as described in Section \ref{sec:obs} on the unsmoothed images.
 The search revealed an X-ray source (with $26.6^{+6.2}_{-5.1}$ and $12.6^{+4.7}_{-3.5}$ soft- and hard-band counts, respectively) $\sim$15\arcsec\ southeast of \hbox{PSS $0926+3055$} as can be seen in Figure~\ref{fig:merged}.
This source is listed in the \textit{NASA/IPAC Extragalactic Database} (NED) as \hbox{CXOGSG J092637.2+305454} and is referenced in \cite{2016ApJS..224...40W} as an \xray\ source, but the number of counts is insufficient for a classification or redshift measurement for this object. 
The source is not detected in any Sloan Digital Sky Survey \citep[SDSS;][]{2017AJ....154...28B} band with upper limits of 23.13, 22.70, and 22.20 mag in the \textit{g', r', \text{and,} i'} bands, respectively ($5\sigma$).
The source is also not detected in observations conducted by the Optical Monitor instrument on-board \xmm\ (\textit{B}-band lower detection limit of 20.7~mag, $5\sigma$) while observing \hbox{PSS $0926+3055$} \citep{2005ApJ...630..729S}, suggesting a red, high-redshift, or highly-obscured object.

Utilizing the \textit{Faint Images of the Radio Sky at Twenty cm} (\textit{FIRST}, \citealt{2015ApJ...801...26H}) survey, the upper limits on the radio flux densities of the 2\arcsec\ regions around \hbox{PSS $0926+3055$} and its potential companion source are $\sim$~0.2~mJy and $\sim0.3$~mJy (1$\sigma$), respectively, at an observed-frame band of 1.4~GHz, where the typically adopted threshold for detection is $\sim$1.0~mJy (5$\sigma$). 

In order to assess the likelihood that the potential companion source is physically associated with our target quasar, we started by computing its mean flux.  
This was performed by using seven of the nine exposures in which the source was significantly detected (it is undetected in the Cycles~19~and~20 exposures).
Following the procedures of Section \ref{sec:obs}, we derived a mean flux for the source in the soft, hard, and full bands, by means of spectral fitting.
We note, however, that since each of the seven exposures did not contain sufficient counts to fit a spectral model, the data from all seven exposures were combined and jointly fit with a single power-law model and a varying intrinsic-absorption model (see Section \ref{sec:meanspectra}).
Using the mean fluxes, we computed the likelihood that one of our four sources would have a companion with at least the mean flux of the potential companion, in the soft, hard, and full band, within a 15\arcsec\ distance.
The mean source flux in each band and the respective likelihoods, $p_c$, to 90\% confidence, are reported in Table \ref{tab:companion}, where $p_c$ is computed using the flux-to-sky-density relation from \citet[][see their Figure 13]{2018MNRAS.478.2132C}. 
The probability of finding a similar or brighter source within 15\arcsec\ of one of our sources is small, although non-negligible, with a mean $p_c$ of 3.1\% across all three bands.

The source's fluxes for each epoch were estimated using \textit{srcflux} with the mean photon index and $N_{H}$ from the combined spectral fit: \hbox{$\Gamma_{0.5-10.0{\rm~keV}}\sim1.6$} and \hbox{$N_{H}\sim2.7\times10^{20} \text{~cm}^{-2}$}.
The source's light curve was then simulated 1000 times (see Section \ref{sec:sims}) and following the $\chi^2$ variability test, the source is not variable in the soft or hard band (\hbox{$p_{s}=0.50$} and \hbox{$p_{s}=0.57$}, respectively).
Follow-up multi-wavelength observations of this source are required for its reliable classification. 

\begin{deluxetable}{lcccr}
\label{tab:companion}
\tablecolumns{5}
\tablecaption{Potential Companion Flux and Likelihood}
\tablehead{
\colhead{Band\tablenotemark{\small \ a}} &
\colhead{} &
\colhead{Flux\tablenotemark{\small b} ($10^{-15}\text{erg s}^{-1}\text{cm}^{-2}$)} &
\colhead{}&
\colhead{$p_c$\tablenotemark{\small \ c}}\\
\cline{1-5}
}
\startdata
Soft & & 7.2 $^{+1.6}_{-1.4}$ & & 2.4$^{+1.0}_{-0.5}$ \\
Hard & & 15.8$^{+3.5}_{-3.1}$ & & 3.5$^{+1.4}_{-1.1}$ \\
Full & & 23.0$^{+5.1}_{-4.4}$ & & 3.1$^{+1.1}_{-0.5}$ \\
\enddata
\tablenotetext{a}{Hard- and full-band upper limits are 10~keV, rather than the 8~keV assumed in this work, for comparison with the results in \cite{2018MNRAS.478.2132C}.}
\tablenotemark{\small b}{Mean flux (90\% confidence) measured with a single power-law fit on a combined spectrum using \texttt{Sherpa}.}
\tablenotetext{c}{The probability of finding a source with at least the flux of \hbox{PSS $0926+3055$}'s potential companion within 15\arcsec\ of any of our sources in units of 10$^{-2}$.}
\end{deluxetable}

\subsection{Mean Spectra} \label{sec:meanspectra}

Our \xray\ observations were designed to provide only the minimum number of counts required for basic time-series analyses, and thus, alone, each observation does not provide a meaningful \xray\ spectrum. However, a combination of all of the \chandra\ observations can give further insight as to the basic X-ray spectral properties of each source, as well as the accuracy of the individual fits used to calculate flux.

\begin{figure*} 
\includegraphics[angle=-90,scale=0.63]{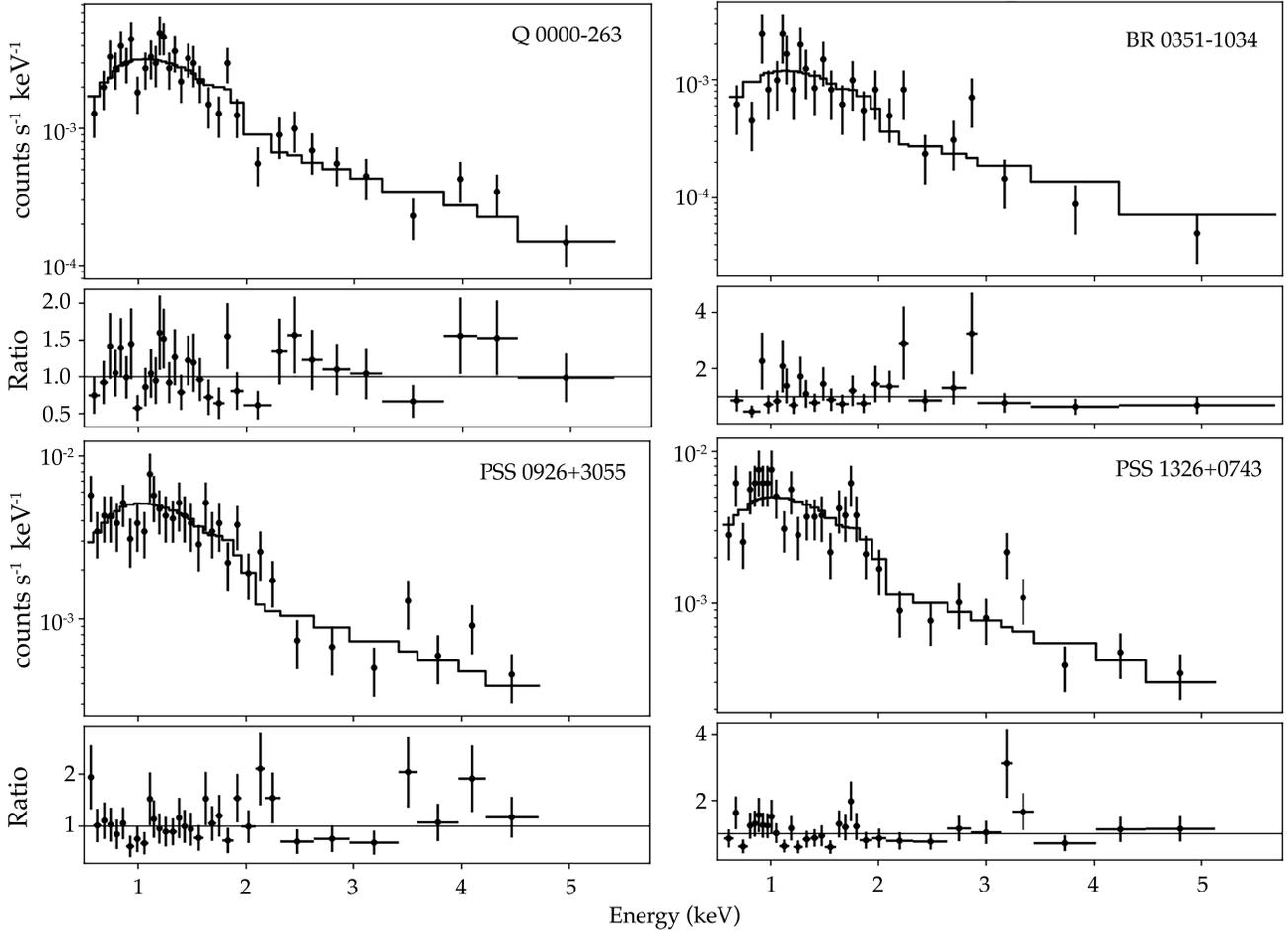}
\caption{\xray\ imaging spectroscopy from each \chandra\ observation combined to produce a mean spectrum for each source. 
Error bars are plotted at 90\% confidence. Solid lines represent our best-fit models with Galactic-absorbed power-law models. Each spectral fit was modeled using the \textit{cstat} statistic on data grouped into a minimum of one count per energy bin. 
Plotted below each spectrum is the ratio of the best-fit model (i.e., data/model).
The spectra were binned for presentation purposes into groups of nine counts, except for our faintest source, \hbox{BR $0351-1034$}, for which the data are binned into groups of five counts.
For all four objects, the photon indices obtained with \chandra\ are consistent, within the errors, with the values obtained from the corresponding \xmm\ observations \citep{2005ApJ...630..729S}, as well as the general type-I AGN population up to $z\sim7$, and are reported in Table \ref{tab:spec}.}
\label{fig:spectra}
\end{figure*}

Our mean spectra were obtained using the \textsc{ciao} thread \textit{combine\textunderscore spectra} to merge the spectrum from each level-2 event file into a mean spectrum for each object, and are presented in Figure \ref{fig:spectra}.
\xray\ spectral fitting was performed using \texttt{Sherpa} to fit a single power-law model with a Galactic-absorption component (\texttt{xsphabs} $\times$ \texttt{xspowerlaw}). 
The photon indices were determined by minimization of the \textit{cstat} statistic with the spectra grouped to a minimum of one count per energy bin.
Ratios are presented as data / model.
\begin{deluxetable}{lcccr} 
\tablecolumns{5}
\tablecaption{Properties of Mean Spectra}
\tablehead{
\colhead{}&
\multicolumn{2}{l}{Flux\tablenotemark{a} ($10^{-15}\text{erg s}^{-1}\text{cm}^{-2}$)}\\
\cline{2-3}
\colhead{Quasar} &
\colhead{$f_{0.5-2\text{~keV}}$} &
\colhead{$f_{2-8\text{~keV}}$}&
\colhead{$\Gamma_{\rm eff}$\tablenotemark{\small a}}&
\colhead{$C_\nu$\tablenotemark{\small b}(dof) / $N_{\rm net}$\tablenotemark{\small c}}
}
\startdata
\object{Q~0000$-$263} &   $24\pm2$ & $24\pm2$ & 1.98 $\pm$ 0.16 & 1.13(154) / 332\\
\object{BR~0351$-$1034} & $9\pm1$ & $11^{+4}_{-3}$ & 1.85 $\pm$ 0.26 & 0.73(96) / 128\\
\object{PSS~0926$+$3055} & $33\pm4$ & $45\pm4$ & 1.79 $\pm$ 0.17& 0.96(164) / 320\\
\object{PSS~1326$+$0743} & $33\pm3$ & $43\pm4$ & 1.81 $\pm$ 0.16 & 0.82(165) / 320\\
\enddata
\tablenotetext{a}{90\% Confidence}
\tablenotetext{b}{The reduced \textit{cstat} statistic ($C/\text{dof}$) as a goodness-of-fit measurement \citep{2017AA...605A..51K}.}
\tablenotetext{c}{Net source counts (soft+hard) over all \chandra\ epochs.}
\label{tab:spec}
\end{deluxetable}

Due to the Cycle-to-Cycle decay in the \xray\ ultrasoft band (see Appendix \ref{apx:a}), our spectra were only fit across the observed-frame energy range \hbox{0.5--8~keV}. 
The photon indices we obtain from the stacked \chandra\ data are consistent, within the 90\% uncertainties, with the values obtained from \xmm\ imaging spectroscopy of the sources using a similar model \citep[fit over 0.2--10~keV;][]{2005ApJ...630..729S} roughly three rest-frame years earlier, and are reported in Table~\ref{tab:spec}.
Also reported in Table~\ref{tab:spec} is a model-based estimate of the mean soft- and hard-band flux for each object, calculated with the model component \texttt{xscflux}, and the net counts ($N_{0.5-2.0\text{~keV}}+N_{2-8\text{~keV}}$) for each source. 
All mean fluxes from the stacked spectra are consistent, within the uncertainties, with the mean fluxes computed from the individual observations in Table \ref{tab:lc_chandra}.
 
Our sources' photon indices are consistent with those of typical type-1 AGNs ($\Gamma\sim1.9$ up to $z\sim7$).
Luminous quasar spectra have been individually and jointly fit up to $z\sim7$, and show relatively consistent behavior \citep[see, e.g.,][]{2005A&A...432...15P,2005AJ....129.2519V,2006ApJ...644...86S,2007ApJ...665.1004J,2017A&A...603A.128N,2019A&A...630A.118V,2021ApJ...908...53W}.

\subsection{Ground-Based Photometry}\label{sec:ground_obs}

With the exception of the Cycle 19 observations of \hbox{Q $0000-263$} and \hbox{BR $0351-1034$}, all the \chandra\ observations were complemented by near-simultaneous optical-UV observations to search for similar behaviors in \xray\ and rest-frame UV variability. 
The ground-based observations were performed at the Tel Aviv University Wise Observatory (WO), utilizing the 1~m (WO~1m), C18~18-inch (WOC18), and C28~28-inch (WOC28) telescopes, the Las Campanas Observatory, using the du Pont 2.5~m telescope (LCO-dP), and the Las Cumbres Observatory using the 1~m telescope (LCO-1m). 
The properties of each detector used are reported in Table \ref{tab:telescopes}.
Depending on availability per night, each observatory used the SDSS $u', g', r', i', \text{ and}\ z'$ filters \citep{1996AJ....111.1748F}, and the Bessel $B, V, R \text{ and}\ I$ filters \citep{1998A&A...333..231B}.

\begin{deluxetable}{ccCC}
\tablecaption{Ground-Based Telescope Properties}
\tablecolumns{4}
\tablehead{
\colhead{Telescope}&
\colhead{Detector (CCD)}&
\colhead{FOV}&
\colhead{Spatial Scale ($\text{arcsec~pix}^{-1}$)}
}
\startdata
WO~1m\tablenotemark{\scriptsize a} 	& 	PI VA1300B			& 13\arcmin\times13\arcmin		 	& 0.58	\\
&			SBIG STX16803			& 17\arcmin.82\times 17\arcmin.82	& 0.78	\\
WOC18\tablenotemark{\scriptsize a} 	&	QSI~683			& 48\arcmin.9\times36\arcmin.8		& 0.88	\\
WOC28\tablenotemark{\scriptsize a} 	&	FLI~PL16801 	& 56\arcmin.9\times56\arcmin.9 		& 0.83 	\\
LCO-dP\tablenotemark{\scriptsize b}	&	Wide~Field		& 25\arcmin\text{~Diameter}			& 0.48 	\\
LCO~1m\tablenotemark{\scriptsize c}&	Sinistro		& 26\arcmin.5\times26\arcmin.5		& 0.39
\enddata
\tablenotetext{a}{
\url{https://physics.tau.ac.il/astrophysics/wise_observatory_manuals}} 
\tablenotetext{b}{\url{http://www.lco.cl/?epkb_post_type_1=wide-field-reimaging-ccd-camera-wfccd}}
\tablenotetext{c}{
\url{https://lco.global/observatory/instruments/sinistro/}}
\label{tab:telescopes}
\end{deluxetable}

The same analysis procedures used in Papers~\hyperlink{P1}{I}~and~\hyperlink{P2}{II} were followed to obtain the calibrated magnitudes and rest-frame UV flux densities, which are reported in Tables \ref{tab:ground} and \ref{tab:lc_chandra_optical}, respectively. 
The light-curve calibration is done on the entire set of observations, rather than each individual image. Therefore, each new observation has the potential of showing a systematic change in magnitudes for all previous observations. 
Between Papers~\hyperlink{P1}{I}~and~\hyperlink{P2}{II}, only two non-negligible, but small (i.e., consistent with the original values at the $\sim2\sigma$ level) changes were reported. 
All values reported in Paper~\hyperlink{P2}{II} are consistent with those calculated after re-calibration in this work, and are left unchanged in Table \ref{tab:ground}.

	\subsubsection{Optical-UV Variability}

To test for variability in the optical-UV bands, the $\chi^{2}$ variability test was applied to the light curve of each band with over three observations, per object, as well as measurement of the respective variability amplitude. 
The 90\% confidence level in the \xray\ analysis was selected due to the large uncertainty in the flux estimates ($\sim$30\%) driven by the low-count photon statistics. 
The optical/UV flux measurements are comparatively more accurate (generally $\lesssim0.1\%$ uncertainty), and we therefore adopt a variability criterion of $p\geq99\%$.

Variability is considered a property of a source's entire light curve, rather than a property of individual events at specific points in the light curve. 
In order to maintain the qualitative nature of a band being variable, the $\chi^2$ and $\sigma^2_{\text{rms}}$ results need to be considered independently, as well as together. 
In some cases, $\sigma^2_{\text{rms}}$ and its uncertainty is inconsistent with zero, suggesting variability, but the $\chi^2$ result does not allow us to reject the null hypothesis confidently.
Similarly, in some other cases, the $\chi^2$ result allows us to reject the non-variable null hypothesis, though the variability amplitude is consistent with zero within its uncertainty. 
Each of these cases were inspected epoch-by-epoch to see if variability was a general trend in the light curve, or if a single event (i.e., a one-time decrease or increase in flux) caused the disagreement between the tests. 
If a single event was responsible, and the remaining epochs produced consistent results in both tests, the band was \textit{not} considered variable. 

The following objects are considered variable, at $> 99\%$ confidence, in the following bands: \hbox{Q $0000-263$} in \textit{g', r', V} and \textit{R}, \hbox{PSS $0926+3055$} in \textit{g', r', V, R,} and \textit{I}, and \hbox{PSS $1326+0743$} in \textit{g'}. 
The first half of epochs in the single band that \hbox{BR $0351-1034$} was determined variable (\textit{g'}) was subject to errors four times larger than the later epochs, which makes it difficult to determine the reliability of the test results: $p>0.99,\ \sigma^2_{\text{rms}}=0.05\pm0.03$. 
With the exception of \hbox{BR $0351-1034$}, the mean variability amplitude of the \chandra\ sources was $<0.01$ in all tested bands. 
In general, the \chandra\ sources appear optically variable at a level of $<0.1$ mag on rest-frame timescales of $\sim$2--3 months.

\subsubsection{UV Flux Density and Optical-to-\xray\ Spectral Slope}
\begin{figure} 
\epsscale{1.4}
\plotone{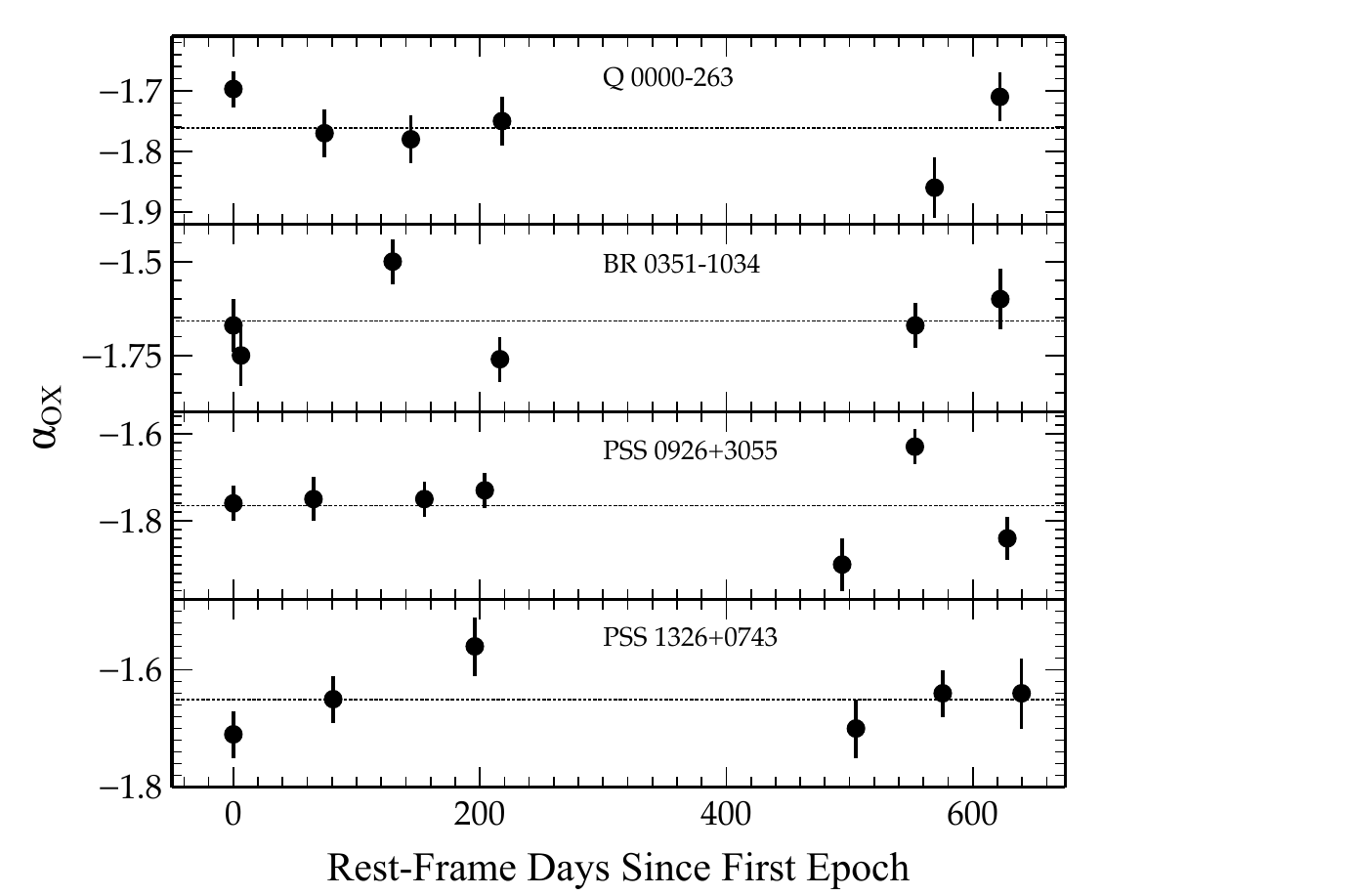}
\caption{Optical-to-\xray\ spectral slope, \aox\, plotted as a function of rest-frame days since the first optical epoch. The dashed line in each panel is the mean \aox\ for the object over all epochs. The large $\alpha_{\textsc{ox}}$ dispersion in the later epochs of \hbox{PSS $0926+3055$} correspond to significant epoch-to-epoch soft-band \xray\ variability, suggesting that scatter in \aox\ is dominated by \xray\ variability.}
\label{fig:a_oxcurves}
\end{figure}

Rest-frame UV flux densities at 1450~\AA\ are reported in Table \ref{tab:lc_chandra_optical} for each ground-based epoch.
The band used to calculate the flux density was chosen to minimize emission-line contamination at the rest-frame wavelength, while providing the smallest difference between the band wavelength and $1450(1+z)$~\AA, and maintaining the highest S/N ratio.
Once a band was selected, the flux density at a rest-frame wavelength of 1450~\AA\ was extrapolated from the flux density at the effective wavelength of the respective band by assuming a quasar continuum of the form \hbox{$f_{\lambda} \propto \lambda^{-1.5}$} (\hbox{corresponding to $f_{\nu} \propto \nu^{-0.5}$}, \citealt{2001AJ....122..549V}).
Magnitude zero-points used in the magnitude-to-flux conversions were obtained from \cite{1996AJ....111.1748F} and \cite{1998A&A...333..231B}. 
With the exception of \hbox{Q $0000-263$} ($p=0.93$), all the sources are variable in $F_{\text{1450\AA}}$ at $>99\%$ confidence.

Extrapolating the rest-frame 2500~\AA\ flux density, along with the rest-frame 2-keV flux density ($f_{\text{2~keV}}$, Table \ref{tab:chandra_counts}, also see Table \ref{tab:apxtable}) and its uncertainty (derived from the soft-band fluxes in Table \ref{tab:lc_chandra}), we compute the optical-to-\xray\ spectral slope, $\alpha_{\textsc{ox}}$, defined as
$\log{(f_{\text{2~keV}}/f_{\text{{2500 \AA}}})}/\log{(\nu_{\text{2~keV}}/\nu_{\text{{2500 \AA}}})}$, and report the results in Table \ref{tab:lc_chandra_optical}.

\begin{deluxetable*}{lllcccccccc}
\tablecolumns{11}
\tablecaption{Ground-Based Photometry}
\tablehead{ 
\colhead{} &
\colhead{} &
\colhead{Obs.} &
\colhead{$g'$} &
\colhead{$r'$} &
\colhead{$i'$} &
\colhead{$z'$} &
\colhead{$B$} &
\colhead{$V$} &
\colhead{$R$} &
\colhead{$I$} \\
\colhead{Quasar} &
\colhead{Obs.\tablenotemark{a}} &
\colhead{Date} &
\colhead{(mag)} &
\colhead{(mag)} &
\colhead{(mag)} &
\colhead{(mag)} &
\colhead{(mag)} &
\colhead{(mag)} &
\colhead{(mag)} &
\colhead{(mag)}
}
\startdata
\object{Q~0000$-$263}
& WO1m   & 2011 Sep 4  & 18.93$\pm$0.02 & 17.45$\pm$0.02 & \nodata & \nodata & 19.58$\pm$0.04 & 18.23$\pm$0.02 & 17.16$\pm$0.02 & \nodata \\
& WO1m   & 2012 Sep 14 & 18.93$\pm$0.03 & 17.48$\pm$0.01 & \nodata & \nodata & 19.45$\pm$0.09 & 18.28$\pm$0.02 & 17.18$\pm$0.03 & \nodata \\
& WO1m   & 2012 Sep 15 & 18.97$\pm$0.02 & 17.48$\pm$0.01 & \nodata & \nodata & 19.53$\pm$0.04 & 18.26$\pm$0.02 & 17.17$\pm$0.01 & \nodata \\
& WOC18 & 2013 Sep 5 & \nodata & \nodata & \nodata & \nodata & 19.62$\pm$0.10 & 18.37$\pm$0.04 &
17.21$\pm$0.02 & \nodata \\
& WOC18 & 2014 Sep 19 & \nodata & \nodata & \nodata & \nodata & \nodata & 18.18$\pm$0.04 &
17.07$\pm$0.03 & \nodata \\
& WOC18 & 2014 Sep 20 & \nodata & \nodata & \nodata & \nodata & 19.40$\pm$0.06 & 18.14$\pm$0.04 &
17.09$\pm$0.02 & \nodata \\
& LCO-dP & 2014 Sep 26 & \nodata & \nodata & \nodata & \nodata & \nodata & 17.97$\pm$0.14 & \nodata & \nodata \\
& WOC28 & 2019 Aug 13 & 18.71$\pm$0.10 & \nodata & \nodata & \nodata & \nodata & \nodata & 16.97$\pm$0.04 & \nodata \\
& LCO-1m & 2020 Aug 21 & 18.68$\pm$0.05 & 17.23$\pm$0.02 & \nodata & \nodata & \nodata & \nodata & \nodata & \nodata \\
\cline{2-11}
%
\object{BR~0351$-$1034}	
& WO1m  & 2011 Mar 3  & \nodata & 19.39$\pm$0.06 & \nodata & \nodata & \nodata & \nodata & 19.24$\pm$0.05 & \nodata    \\
& WO1m  & 2011 Mar 5  & \nodata & 19.33$\pm$0.04 &  \nodata & \nodata & \nodata & \nodata & \nodata &\nodata \\
& WO1m  & 2011 Sep 26 & \nodata & 19.33$\pm$0.03 & \nodata & \nodata & \nodata & 20.59$\pm$0.09 & 19.29$\pm$0.04 & \nodata\\
& LCO-dP & 2011 Oct 29 & \nodata & \nodata & \nodata & \nodata & 22.79$\pm$0.11 & 20.55$\pm$0.02 & 19.35$\pm$0.03 & \nodata \\
& LCO-dP  & 2013 Aug 4 & \nodata & \nodata & \nodata & \nodata & 22.46$\pm$0.06 & 20.36$\pm$0.04 & \nodata & \nodata \\
& WO1m  & 2013 Aug 18 & \nodata & \nodata & \nodata & \nodata & \nodata & 20.39$\pm$0.09 & 19.23$\pm$0.08 & \nodata \\
& WO1m  & 2014 Nov 25 & \nodata & \nodata & \nodata & \nodata & \nodata & 20.54$\pm$0.11 & 19.14$\pm$0.06 & \nodata \\
& LCO-dP & 2014 Nov 26 & \nodata & \nodata & \nodata & \nodata & \nodata & 20.41$\pm$0.04 & 19.10$\pm$0.04 & \nodata \\
& WO1m & 2014 Nov 26 & \nodata & \nodata & \nodata & \nodata &  22.51$\pm$0.08 & \nodata & \nodata & \nodata \\
& WOC18 & 2019 Nov 2 & 20.38$\pm$0.31 & 18.98$\pm$0.12 & \nodata & \nodata & \nodata & \nodata & \nodata & \nodata \\
& WOC18 & 2019 Nov 3 & 20.34$\pm$0.26 & 19.26$\pm$0.11 & \nodata & \nodata & \nodata & \nodata & \nodata & \nodata \\
& WOC28 & 2019 Nov 19 & \nodata & 19.37$\pm$0.08 & \nodata & \nodata & \nodata & \nodata & \nodata & \nodata \\
& LCO-1m & 2020 Nov 6 & 20.93$\pm$0.07 & 19.18$\pm$0.03 & \nodata & \nodata & \nodata & \nodata & \nodata & \nodata \\
& LCO-1m & 2020 Nov 9 & 21.16$\pm$0.08 & \nodata & \nodata & \nodata & \nodata & \nodata & \nodata & \nodata \\
\cline{2-11}
%
\object{PSS~0926$+$3055} 
& WO1m  & 2011 Mar 4 & 18.45$\pm$0.01 & 17.13$\pm$0.01 & 17.01$\pm$0.01 & 17.22$\pm$0.03 & 
\nodata & 17.83$\pm$0.02 & 16.90$\pm$0.01 & 16.60$\pm$0.02 \\
& WO1m  & 2012 Feb 4 & 18.55$\pm$0.05 & 17.23$\pm$0.04 & 17.05$\pm$0.05 & \nodata & 
\nodata & 17.94$\pm$0.05 & 17.11$\pm$0.08 & 16.66$\pm$0.04 \\
& WO1m  & 2013 May 15 & \nodata & \nodata & \nodata & \nodata & 
19.20$\pm$0.07 & 17.91$\pm$0.01 & 16.92$\pm$0.01 & 16.58$\pm$0.01 \\
& WO1m  & 2014 Jan 23  & 18.43$\pm$0.03 & 17.13$\pm$0.02 & 17.00$\pm$0.02 & \nodata & 
\nodata & 17.91$\pm$0.03 & 16.91$\pm$0.02 & 16.41$\pm$0.02 \\
& WO1m & 2018 Mar 9 & \nodata & 17.36$\pm$0.01 & 17.13$\pm$0.01 & 17.14$\pm$0.13 & 18.14$\pm$0.01 & 18.14$\pm$0.01 & 17.09$\pm$0.06 & 16.76$\pm$0.06 \\
& WO1m\tablenotemark{b} & 2018 Mar 11 & \nodata & \nodata & \nodata & 17.18$\pm$0.13 & 19.43$\pm$0.09 & 18.08$\pm$0.08 & 17.07$\pm$0.08 & 16.69$\pm$0.06 \\
& WOC28 & 2019 Jan 10 & \nodata & 17.18$\pm$0.05 & 17.02$\pm$0.03 & 17.07$\pm$0.06 & \nodata & \nodata & \nodata & \nodata \\
& LCO-1m & 2020 Feb 2 & 18.56$\pm$0.01 & 17.15$\pm$0.03 & 17.04$\pm$0.01 & 17.06$\pm$0.01 & \nodata & \nodata & \nodata & \nodata \\
\cline{2-11}
%
\object{PSS~1326$+$0743}
& WO1m  & 2011 Mar 8  & 19.15$\pm$0.10 & \nodata & \nodata & \nodata & \nodata & 18.47$\pm$0.03 & 17.48$\pm$0.02 & 16.88$\pm$0.03   \\
& WO1m  & 2011 Mar 14 & 19.28$\pm$0.03 & 17.82$\pm$0.10 & 17.51$\pm$0.10 & 17.15$\pm$0.03 & 
\nodata & 18.47$\pm$0.02 & 17.49$\pm$0.02 & 16.77$\pm$0.12   \\
& WO1m  & 2012 May 1  &        \nodata & 17.79$\pm$0.06 & 17.61$\pm$0.07 &        \nodata & 
\nodata & 18.52$\pm$0.14 & 17.59$\pm$0.07 & 16.69$\pm$0.09 \\  
& WO1m  & 2013 Dec 15  &  19.46$\pm$0.12 & 17.81$\pm$0.03 & 17.61$\pm$0.09 & \nodata & 
\nodata & 18.64$\pm$0.10 & 17.54$\pm$0.04 & 16.96$\pm$0.10 \\  
& WO1m  & 2013 Dec 16  &  19.25$\pm$0.06 & 17.80$\pm$0.02 & 17.60$\pm$0.04 & \nodata & 
20.07$\pm$0.20 & 18.66$\pm$0.06 & 17.53$\pm$0.02 & 16.90$\pm$0.03 \\
& WO1m  & 2018 May 2  &  \nodata & \nodata & \nodata & \nodata & \nodata & 18.72$\pm$0.16 & 17.60$\pm$0.07 & 16.97$\pm$0.08 \\
& WOC28  & 2019 May 1  &  19.34$\pm$0.06 & \nodata & \nodata & \nodata & \nodata & \nodata & 17.54$\pm$0.04 & \nodata \\
& WOC28  & 2019 May 4  &  19.40$\pm$0.04 & \nodata & \nodata & \nodata & \nodata & \nodata & 17.57$\pm$0.02 & \nodata \\
& WOC28  & 2019 May 5  &  19.38$\pm$0.05 & \nodata & \nodata & \nodata & \nodata & \nodata & 17.57$\pm$0.03 & \nodata \\
& LCO-1m  & 2020 Mar 27  & 19.59$\pm$0.02 & 17.86$\pm$0.01 & 17.63$\pm$0.01 & 17.34$\pm$0.01 & \nodata & \nodata & \nodata & \nodata \\
\enddata
\label{tab:ground}
\tablenotetext{a}{Unless otherwise noted, observations made after 2014 with the WO1m utilized the SBIG~STX16803 camera rather than the PI~VA1300B camera. See Section \ref{sec:ground_obs} for details on the detectors.} 
\tablenotetext{b}{Used the PI~VA1300B camera.}
\end{deluxetable*}
\begin{deluxetable*}{llclccc}
\tablecolumns{7}
\tablecaption{Rest-Frame UV Flux Densities and \aox\ Data for the \chandra\ Sources}
\tablehead{
\colhead{Quasar} &
\colhead{JD} &
\colhead{$F_{\lambda}$\tablenotemark{a}} &
\colhead{Obs.} &
\colhead{Band} &
\colhead{\aox\tablenotemark{b}} &
\colhead{$\Delta t$\tablenotemark{c}}
}
\startdata
\object{Q~0000$-$263}
& 2455809.5 & 2.41$\pm$0.04 & WO1m & $R$ & $-1.70\pm0.03$ & $1.4$ \\
& 2456185.5 & 2.35$\pm$0.07 & WO1m  & $R$ & $-1.77\pm0.04$ & $2.4$ \\
& 2456186.5 & 2.39$\pm$0.03 & WO1m  & $R$ & \nodata & \nodata \\
& 2456541.5 & 2.29$\pm$0.05 & WOC18 & $R$ & $-1.78\pm0.04$ & $0.2$ \\
& 2456920.5 & 2.62$\pm$0.08 & WOC18 & $R$ & $-1.75\pm0.04$ & $0.7$ \\
& 2456921.5 & 2.57$\pm$0.05 & WOC18 & $R$ & \nodata & \nodata \\
& 2456927.7 & 1.49$\pm$0.19 & LCO-dP   & $V$ & \nodata & \nodata \\
& 2458709.5 & 2.86$\pm$0.11 & WOC28 & $R$ & $-1.86\pm0.05$ & $0.2$ \\
& 2459038.6 & 2.78$\pm$0.05 & LCO-1m & $r'$& $-1.71\pm0.04$ & $1.4$ \\
\cline{2-7}
\object{BR~0351$-$1034} 
& 2455624.2  & 0.33$\pm$0.02 & WO1m & $R$ & \nodata & \nodata \\
& 2455626.2  & 0.37$\pm$0.01 & WO1m & $r'$ & \nodata & \nodata \\
& 2455831.5  & 0.31$\pm$0.01 & WO1m & $R$ & $-1.65\pm0.07$ & $0.7$\\
& 2455864.8  & 0.30$\pm$0.01 & LCO-dP & $R$ & $-1.75\pm0.08$ & $0.4$ \\
& 2456509.8  & 0.15$\pm$0.01 & LCO-dP & $V$ & \nodata & \nodata \\
& 2456523.5  & 0.33$\pm$0.03 & WO1m & $R$ & $-1.50\pm0.06$ & $6.0$ \\
& 2456987.5  & 0.36$\pm$0.02 & WO1m & $R$ & \nodata & \nodata \\
& 2456988.5  & 0.37$\pm$0.01 & LCO-dP & $R$ & $-1.76\pm0.06$ & $0.2$ \\
& 2458790.5  & 0.52$\pm$0.06 & WOC18 & $r'$ & \nodata & \nodata \\
& 2458791.5  & 0.40$\pm$0.04 & WOC18 & $r'$ & $-1.67\pm0.06$ & $<0.1$ \\
& 2458806.4  & 0.36$\pm$0.03 & WOC28 & $r'$ & \nodata & \nodata \\ 
& 2459160.8  & 0.43$\pm$0.12 & LCO-1m & $r'$ & $-1.60\pm0.08$ & $1.8$ \\
\cline{2-7}
\object{PSS~0926$+$3055} 
& 2455625.2 & 2.81$\pm$0.06 & WO1m & $I$ & $-1.76\pm0.04$ & $0.3$ \\
& 2455962.3 & 2.68$\pm$0.11 & WO1m & $I$ & $-1.75\pm0.05$ & $4.4$ \\
& 2456428.5 & 2.87$\pm$0.03 & WO1m & $I$ & $-1.75\pm0.04$ & $0.8$ \\
& 2456681.5 & 3.36$\pm$0.07 & WO1m & $I$ & $-1.73\pm0.04$ & $1.2$ \\
& 2458187.4 & 2.50$\pm$0.02 & WO1m & $I$ & $-1.90\pm0.06$ & $0.3$ \\
& 2458189.3 & 2.59$\pm$0.14 & WOC18 & $I$ & \nodata & \nodata \\
& 2458494.4 & 2.96$\pm$0.08 & WOC28 & $i'$ & $-1.63\pm0.04$ & $0.6$ \\
& 2458882.8 & 2.91$\pm$0.03 & LCO-1m & $i'$ & $-1.84\pm0.05$ & $1.8$ \\
\cline{2-7}
\object{PSS~1326$+$0743} 
& 2455629.6 & 1.76$\pm$0.04 & WO1m & $R$ & $-1.71\pm0.04$ & $0.4$ \\
& 2455635.5 & 1.74$\pm$0.03 & WO1m  & $R$ & \nodata & \nodata \\
& 2456049.3 & 1.59$\pm$0.11 & WO1m  & $R$ & $-1.65\pm0.04$ & $0.3$ \\
& 2456642.5 & 1.65$\pm$0.06 & WO1m  & $R$ & $-1.56\pm0.05$ & $1.9$ \\
& 2458241.3 & 1.57$\pm$0.10 & WO1m & $R$ & $-1.70\pm0.05$ & $0.5$ \\
& 2458605.4 & 1.66$\pm$0.61 & WOC28 & $R$ & $-1.64\pm0.04$ & $0.4$ \\
& 2458608.4 & 1.16$\pm$0.02 & WOC28 & $R$ & \nodata & \nodata \\
& 2458609.3 & 1.16$\pm$0.03 & WOC28 & $R$ & \nodata & \nodata \\
& 2458936.2 & 1.70$\pm$0.02 & LCO-1m & $i'$ & $-1.64\pm0.06$ & $0.3$
\enddata
\tablecomments{For each source, \aox\ is given only for the shortest time separations between the optical and \chandra\ observations.}
\tablenotetext{a}{Flux density at rest-frame 1450~\AA\ in units of $10^{-16}$~erg~cm$^{-2}$~s$^{-1}$~\AA$^{-1}$, extrapolated from
the flux density at the effective wavelength of the respective band, assuming a continuum of the form
\hbox{$f_{\lambda} \propto \lambda^{-1.5}$} (\hbox{corresponding to $f_{\nu} \propto \nu^{-0.5}$}, \citealt{2001AJ....122..549V}).}
\tablenotetext{b}{Errors at the 1-$\sigma$ level on \aox\ were derived according to \S~1.7.3 of \citet{1991pgda.book.....L}, given the errors on the
rest-frame UV flux densities and the errors on the \xray\ fluxes from Table~\ref{tab:lc_chandra}.}
\tablenotetext{c}{Rest-frame days between the ground-based and \chandra\ observations.}
\label{tab:lc_chandra_optical}
\end{deluxetable*}
Also reported in Table \ref{tab:lc_chandra_optical} are the rest-frame time separations, $\Delta t$, between the \xray\ and corresponding optical-UV observations.
These time separations were kept as small as possible, and are all on the order of $\approx1$~day in the rest frame. 
The \aox\ measurements are also presented in Figure~\ref{fig:a_oxcurves} as a function of rest-frame days since the first optical observation.
Paper~\hyperlink{P2}{II} reported a significant change in $\alpha_{\textsc{ox}}$ at a level of \hbox{$\Delta\alpha_{\textsc{ox}}=0.08\ \text{and}\ \Delta\alpha_{\textsc{ox}}=0.09$} between Cycle pairs for \hbox{Q $0000-263$} and \hbox{PSS $0926+3055$}, respectively.
After \xray\ fluxes were estimated using \texttt{Sherpa}, and three new epochs were added, the previous results have changed for all sources as explained below.

The measurement corresponding to the Cycle-20 \hbox{Q $0000-263$} observation is inconsistent with every other measurement, with the exception of the Cycle-14 measurement. 
The Cycle-20 \chandra\ observation exhibited a decrease of $\sim$32\% ($\sim$52\% at the lower limit) from the mean flux of this source, while the optical observation, $\sim$5 rest-frame hours later, deviated $<$1\% from the mean flux. 
The maximum variation is between Cycles 12 and 20 with $\Delta\alpha_{\textsc{ox}} = 0.17$.
BR $0351-1034$ shows significant variations up to \hbox{$\Delta\alpha_{\textsc{ox}} = 0.26$}, however, the source's low $S/N$ should be considered when interpreting this result. 

PSS $0926+3055$ showed significantly consistent measurements until Cycle 19, which was an extreme in its \xray\ light curve with a flux 57\% lower than the mean.
The Cycle-20 \xray\ flux was 57\% \textit{higher} than the mean.
Between Cycle pairs 19--20, and 20--21, we find \hbox{$\Delta\alpha_{\textsc{ox}} = 0.27$} and \hbox{$\Delta\alpha_{\textsc{ox}} = 0.21$}, respectively.
In \hbox{PSS $1326+0743$}, $\alpha_{\textsc{ox}}$ shows no significant variability and the largest variation is \hbox{$\Delta\alpha_{\textsc{ox}} = 0.04$}, which corresponds to the Cycle-14 flux density estimate rising 62\% above the mean 

We find that variations in photon indices and \xray\ flux produce most of the variation in $\alpha_{\textsc{ox}}$, which supports findings that scatter in $\alpha_{\textsc{ox}}$ is driven by \xray\ variability (see, e.g., \hyperlink{P1}{Paper~I} and references therein).
However, the $\chi^{2}$ test shows that \aox\ is not considered variable overall in each source (90\% confidence), with the exception of \hbox{PSS $0926+3055$} ($p=0.93$).

To further characterize the role of \xray\ variability in the \aox\ scatter, we computed \aox\ with a constant photon index for each source across all epochs (i.e., with the \textit{srcflux} fluxes).
In this case, all four sources were formally non-variable in \aox\ ($\langle p\rangle=0.17$); however, we note that between the Cycle~13 and 14 observations of \hbox{PSS $0926+3055$}, \aox\ showed a significant, yet small, variation at a level of \hbox{$\Delta\alpha_{\textsc{ox}} = 0.05$}.
These results suggest that if the photon index remains unchanged, changes in \aox\ are driven by \xray\ flux variability.

The mild variations in \aox\ observed for our sources are consistent with observations of luminous, high-redshift quasars.
For example, in an effort to tightly constrain $\alpha_{\textsc{ox}}$ dispersion to use quasars as cosmological standard candles, \cite{2018A&A...619A..95C} showed that RQQs up to $z\sim3.5$ show $\langle\Delta\alpha_{\textsc{ox}}\rangle \sim0.08$ on 10--100-day timescales (uncorrected for photometric uncertainty).
Our sources behave consistently, on average, with their results with $\langle\Delta\alpha_{\textsc{ox}}\rangle \sim0.09\pm0.01$. 
While our results do not suggest increased $\alpha_{\textsc{ox}}$ variability with redshift, the \cite{2018A&A...619A..95C} results, that are based on a much larger sample, leave a dispersion of $\Delta\alpha_{\textsc{ox}}\sim0.2$ unaccounted for by intrinsic source variability, and a redshift dependence is not ruled out \citep[see, also, ][]{2016ApJ...819..154L}.

\section{Summary} \label{sec:summary}

We present three new \xray\ epochs in a long-term time-series analysis of four luminous high-redshift \hbox{($z\approx4.1 - 4.4$)} RQQs extending our temporal baseline to $\sim$1300 days in the rest frame for half of our sources and $\sim$2000 days in the rest frame for the other half.
Our new \xray\ observations were obtained with \chandra\ and are accompanied by near-simultaneous ground-based optical-UV observations. Our findings can be summarized as follows:
\begin{enumerate}
	\item We find that two of our four sources, \hbox{Q $0000-263$} and \hbox{PSS $0926+3055$}, are \xray\ variable in the soft band and three sources, \hbox{BR $0351-1034$}, \hbox{PSS $0926+3055$}, and \hbox{PSS $1326+0743$}, are variable in the hard band.
    \item There is no evidence for increased \xray\ variability as redshift increases, bolstering our findings from \hyperlink{cite.2017ApJ...848...46S}{Paper~II}. Hard-band \xray\ variability is consistent with that of the soft band.
    \item We do not observe significant changes in the SF of any of our \chandra\ sources across all timescales probed, and the ensemble SF of all four sources shows no indication for increased \xray\ variability as redshift increases.
    At all timescales, the soft-band ensemble SF is consistent with or lower than that of the similarly luminous \swift\ sources. The hard-band ensemble SF is consistent with the \swift\ sources at longer timescales, but is highly uncertain at shorter timescales.
    \item The stacked \chandra\ image of \hbox{PSS $0926+3055$} reveals a non-variable \xray\ point source $\sim$15\arcsec\ southeast of the target quasar. While present in a \chandra\ point source catalog, there is currently no record of this source in any other catalog. Additional multi-wavelength observations, aimed particularly at measuring the redshift of this source, are required to determine if it is a companion of \hbox{PSS $0926+3055$}.
    \item Mean photon indices of our sources, measured from their stacked \chandra\ spectra, are consistent, within the uncertainties, with those obtained $\sim$3 rest-frame years prior with \xmm, and with the mean photon index of AGNs observed to $z\sim7$.
    \item All four sources showed epoch-to-epoch variations in $\alpha_{\textsc{ox}}$ on the order of $\langle\Delta\alpha_{\textsc{ox}}\rangle \sim0.09\pm0.01$, however, only \hbox{PSS $0926+3055$} is formally considered variable in $\alpha_{\textsc{ox}}$ over its entire light curve. Variations in \aox\ are contemporaneous with the \xray\ variability, supporting the suggestion that \xray\ variability dominates the scatter in the optical-to-\xray\ spectral slope. Overall, \aox\ variations are consistent with those of RQQs at lower redshifts. 
\end{enumerate}

We aim at continuing this monitoring project in order to (1) extend the temporal baseline, (2) provide the longest term and most thorough \xray\ time-series analysis, and (3) explore short-term behavior, i.e., hourly--daily timescales in the rest frame of RQQs at $z\sim4$; we will also increase the exposure times in future observations given the loss of sensitivity of the ACIS detector (see Appendix \ref{apx:a}).
The \chandra\ monitoring project will be complementary to the \textit{eROSITA} survey \citep{2014AN....335..517P}, which has the capability to detect sources at $z>4$, but may not provide light curves with sufficiently high signal-to-noise ratios for meaningful time-series analyses of such sources. \\

The scientific results presented in this paper are based on observations made by the \textit{Chandra \xray\ Observatory} and on data obtained from the \chandra\ Data Archive. 
Support for this work was provided by the National Aeronautics and Space Administration (NASA) through \chandra\ Award Number \hbox{GO8-19072X} (M.T., O.S.) issued by the \textit{Chandra \xray\ Observatory} Center (CXC), which is operated by the Smithsonian Astrophysical Observatory for and on behalf of NASA under contract \hbox{NAS8-03060}.
W.N.B. acknowledges support from CXC grant GO0-21080X and the V.M. Willaman Endowment. 
We thank an anonymous referee for helping us improve this manuscript.
This work is based, in part, on observations obtained with the Tel Aviv University Wise Observatory 1~m, C18, and C28 telescopes.
This research has made use of the NASA/IPAC Extragalactic Database (NED) which is operated by the Jet Propulsion Laboratory, California Institute of Technology, under contract with the National Aeronautics and Space Administration. This research has also made use of data provided by the High Energy Astrophysics Science Archive Research Center (HEASARC), which is a service of the Astrophysics Science Division at NASA/GSFC and the High Energy Astrophysics Division of the Smithsonian Astrophysical Observatory.

\clearpage

\begin{appendix}
\section{Reduced Sensitivity in the ACIS Soft Band}
\label{apx:a}
\restartappendixnumbering
\cite{2000ApJ...534L.139T} reported on damage to the Advanced CCD Imaging Spectrometer (ACIS) by a bombardment of charged particles in Earth's radiation belts, resulting in increased charge transfer inefficiency (CTI) and causing quantum efficiency (QE) to vary with photon energy and the position on the detector.
Alongside this CTI is an additional effect of contamination buildup on the CCDs\footnote{For additional details, see \url{http://hea-www.harvard.edu/~alexey/acis/memos/cont_spat.pdf}}, which the \chandra\ \xray\ Center reports to have markedly increased in 2009\footnote{\url{http://cxc.harvard.edu/ciao/why/acisqecontamN0010.html}}. 
The rapid increase in contaminant has reduced QE even further and limited response to ultrasoft and soft-band photon energies.

\begin{figure}[h!]
    \centering
    \includegraphics[width=9cm]{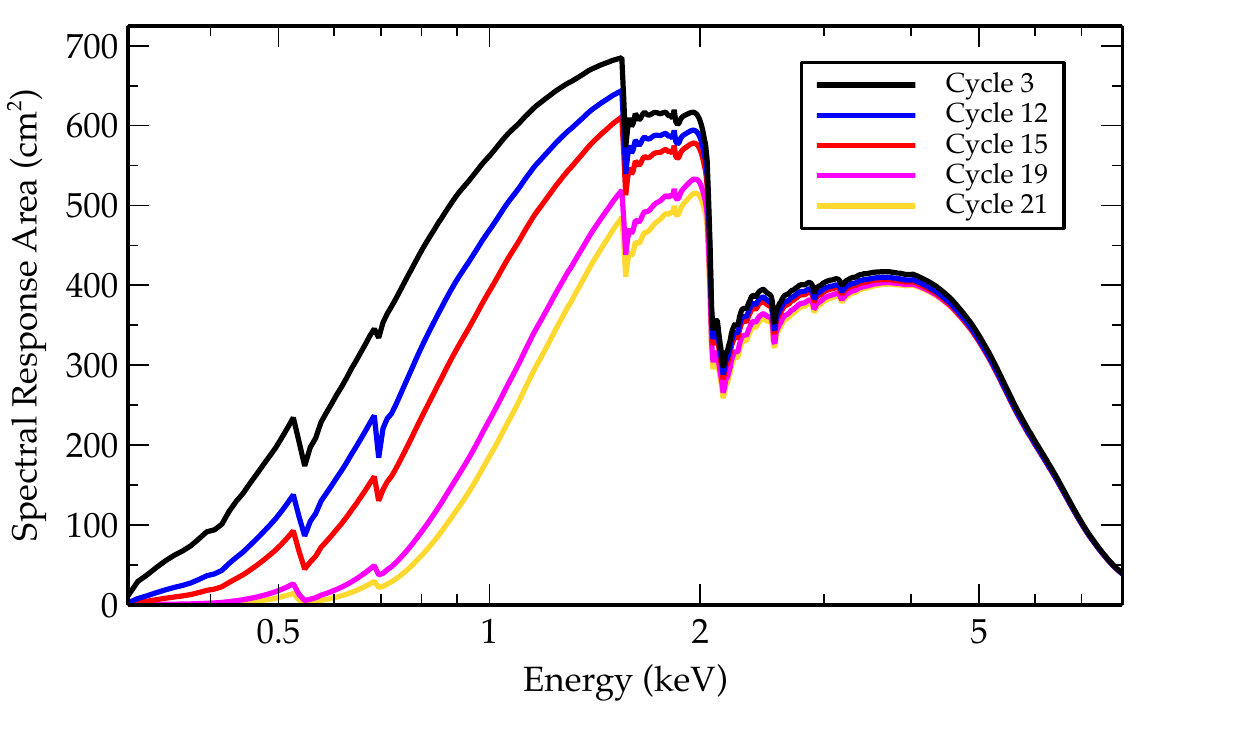}
    \caption{Effective response area of the ACIS-S detector is plotted against energy over select observations of \hbox{PSS $0926+3055$} in our monitoring program. The detector is continually losing effective area at all energy levels (loss at energies $>5$~keV can be seen in Figure \ref{fig:ARFratio}). Detector response below 2~keV is degrading faster than higher energies, and is now nearly zero below 0.5~keV.}
    \label{fig:ARF}
\end{figure}

\begin{figure}[H]
    \centering
    \includegraphics[width=9cm]{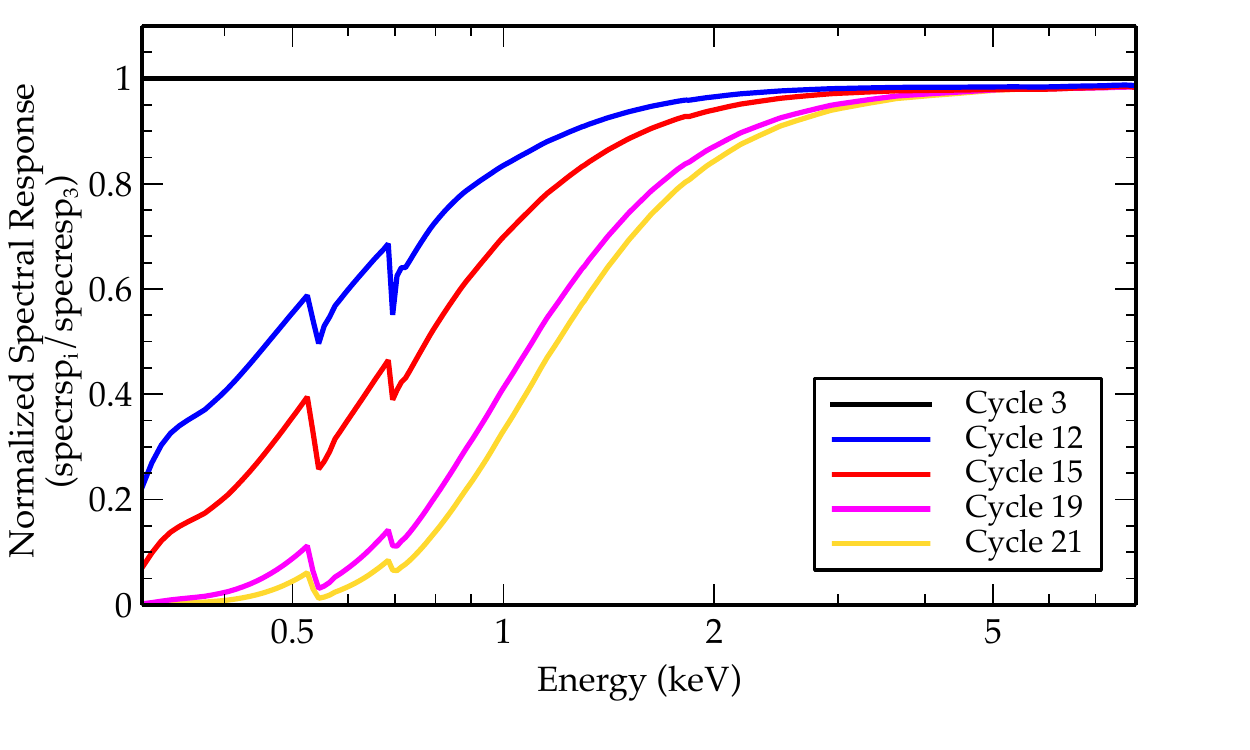}
    \caption{Effective response area, normalized to the first Cycle in our monitoring program, of the ACIS-S detector is plotted against energy over select observations of \hbox{PSS $0926+3055$} in our monitoring program. At all energies, detector response is reduced, and at energies $\lesssim 0.5$~keV, detector response is less than 5\% of that in Cycle 3.}
    \label{fig:ARFratio}
\end{figure}

All data reduction using \textsc{ciao} accounts for the CTI with \textsc{caldb}, which is updated regularly throughout the \chandra\ Cycles  (i.e., years), but there is still a loss of information with the instrument being less sensitive to softer photons. 
The detector's spectral response areas (i.e., ``effective'' area accounting for QE) for select observations of \hbox{PSS $0926+3055$} in the monitoring program are presented in \hbox{Figure \ref{fig:ARF}}. It can be seen that detector response is affected at all energies, and, below $\sim2$~keV, is reduced considerably.
 In \hbox{Figure \ref{fig:ARFratio}}, the same detector response areas are normalized to that of the first Cycle in our monitoring program. 
 The response in the ultrasoft band is now less than $5\%$ of that in Cycle 3, and the response in the soft band ranges from 5\% at 0.5~keV to 80\% at 2~keV of that in Cycle 3.

Effects of this loss of sensitivity can be seen in \hbox{Figure \ref{fig:SBlightcurves}}, which is composed of fluxes given in \hbox{Table \ref{tab:lc_chandra}}.
For each object, the exact same instrument, CCDs, and exposure times were used across all epochs (with the exception of the two $\sim$6-ks Cycle-3 observations which used an additional CCD on the detector).
There is a noticeable increase in the size of the error bars for the soft-band flux (left) from Cycle 3 in 2003 to Cycle 21 in 2019--2020, whereas the error bars for the hard-band flux remain relatively consistent. 
Papers~I~and~II reported counts in the ultrasoft band for each object and epoch, but in recent Cycles, the ultrasoft sensitivity is effectively reduced to zero.
An upper limit of three counts in the ultrasoft band 
was recorded for each object in Cycles 19--21, while there were detected counts in the majority of all previously reported epochs. 

To illustrate the practical difference in calibration and soft-photon response, we used \chandra~\textsc{pimms} to estimate the flux of a Cycle-3 observation with a soft-band count rate of 7.5$\times10^{-3}$~photons s$^{-1}$.
We then used the same count rate to predict the flux in Cycle 21, our latest observation Cycle. 
The Cycle-21 flux estimate returned was three times that of Cycle 3, given the same count rate and model parameters.

\section{Reprocessed Epochs From Papers I and II}\label{apx:b}

The previous \chandra\ observations reported in Papers~I~and~II utilized \chandra~\textsc{pimms} to estimate source fluxes, effective photon indices ($\Gamma_{\rm eff}$), and rest-frame 2~keV Galactic-absorption corrected flux densities.
This tool is intended for observation \textit{planning} and should not be used on observed data due to outdated instrument calibration predictions.
Given the QE issues detailed above, the potential effect on the calibration of flux in a time-series analysis over the lifetime of the instrument is not negligible. 
See Section \ref{sec:obs} for further details. 

To obviate consideration of the potential offsets, we opted to calculate source fluxes by spectral modeling of the individual observations using \texttt{XSPEC} models in \texttt{Sherpa} v14.3.
Each observation was fit over 0.5--8.0~keV with a single power-law model convolved with a photoelectric absorption model (\texttt{xsphabs*xspowerlaw}) using the \textit{cstat} statistic on data grouped to a minimum of one count per energy bin.
Then, the models were frozen and convolved with the \texttt{xscflux} model to calculate energy flux.
Using the soft-band flux and $\Gamma_{\rm eff}$, we used the \pimms\ command-line tool to estimate the rest-frame 2-keV flux density.
The effective photon index and flux density for each observation reported in \pIandII\ are reported in Table \ref{tab:apxtable}.

\begin{deluxetable}{lccc}\label{tab:apxtable}
\tablecolumns{4}
\tablecaption{Spectral \xray\ Measurements}
\tablehead{
\colhead{Object}&
\colhead{Cycle}&
\colhead{$\Gamma_{\rm eff}$\tablenotemark{\small a}}&
\colhead{\ $f_{2{\rm~keV}}$\tablenotemark{\small a}}
}
\startdata
\object{Q $0000-263$}
& 12 			& $2.1\pm0.4$			& 2.18 \\
& 13 			& $1.8\pm0.4$ 			& 1.38 \\
& 14 			& $1.9_{-0.5}^{+0.6}$ 	& 1.30 \\
& 15 			& $1.9_{-0.4}^{+0.5}$ 	& 1.70 \\
\object{BR $0351-1034$}
& 12 			& $1.8\pm1.0$ 			& 0.37 \\
& 13 			& $1.5_{-0.8}^{+0.9}$ 	& 0.21 \\
& 14 			& $2.1_{-0.6}^{+0.7}$ 	& 1.05 \\
& 15 			& $1.1_\pm0.8$ 			& 0.26 \\
\object{PSS $0926+3055$}
& \phantom{1}3  & $1.7\pm0.4$ 			& 1.90 \\
& 12 			& $1.7_{-0.4}^{+0.5}$ 	& 1.78 \\
& 13 			& $2.0\pm0.6$ 			& 1.86 \\
& 14 			& $1.6\pm0.4$ 			& 1.95 \\
& 15 			& $1.8\pm0.5$ 			& 2.60 \\
\object{PSS $1326+0734$}
& \phantom{1}3  & $1.7\pm0.4$ 			& 1.62 \\
& 12 			& $1.6\pm0.5$ 			& 1.52 \\
& 13 			& $1.7\pm0.5$ 			& 1.93 \\
& 14 			& $2.4\pm0.6$ 			& 3.44 \\
& 15 			& $2.1_{-0.4}^{+0.5}$ 	& 2.21 \\
\enddata
\tablenotetext{a}{Effective photon indices (0.5--8.0~keV; 90\% confidence) were estimated by spectral modeling in \texttt{Sherpa}. See Section \ref{sec:obs} for details on the assumed model.
Galactic absorption-corrected flux density at rest-frame 2~keV in units of $10^{-31}$~erg~cm$^{-2}$~s$^{-1}$~Hz$^{-1}$ was estimated with the \pimms\ command-line tool v4.11b \citep{1993Legac...3...21M} by extrapolating the respective soft-band flux (see Table \ref{tab:lc_chandra}) and $\Gamma_{\rm eff}$.}

\end{deluxetable}

\end{appendix}
\clearpage
\bibliography{thomas_variability}{}
\bibliographystyle{aasjournal}

\end{document}